\documentclass[fleqn,usenatbib]{mnras}
\usepackage{newtxtext,newtxmath}
\usepackage[T1]{fontenc}
\usepackage{ae,aecompl}

\usepackage{graphicx}	
\usepackage{amsmath}	
\usepackage{amssymb}	
\usepackage[all]{hypcap} 
\usepackage{upgreek}	

\usepackage{etoolbox}
\makeatletter
\patchcmd\@combinedblfloats{\box\@outputbox}{\unvbox\@outputbox}{}{%
   \errmessage{\noexpand\@combinedblfloats could not be patched}%
}%
 \makeatother

\newcommand{\vir}[1]{``#1''}
\renewcommand{\eqref}[1]{equation\ (\ref{#1})}
\newcommand {\kms} {\,{\rm km\,s}^{-1}}
\newcommand {\kpc} {\,{\rm kpc}}
\newcommand{\hi}{\ifmmode{\rm HI}\else{H\/{\sc i}}\fi} 
\newcommand{\de}{\ifmmode{^\circ}\else{$^\circ$}\fi} 
\newcommand{\vlos}{\ifmmode{V_\mathrm{los}}\else{$V_\mathrm{los}$}\fi} 
\newcommand{\vsys}{\ifmmode{V_\mathrm{sys}}\else{$V_\mathrm{sys}$}\fi} 
\newcommand{\vrot}{\ifmmode{V_\mathrm{rot}}\else{$V_\mathrm{rot}$}\fi} 
\newcommand{\vtra}{\ifmmode{V_\mathrm{t}}\else{$V_\mathrm{t}$}\fi} 
\newcommand{\didt}{\ifmmode{\partial i/\partial t}\else{$\partial i/\partial t$}\fi}
\newcommand {\mo}{{\rm M}_\odot}

\newcommand {\pc} {\,{\rm pc}}
\newcommand {\vdisp}{{\sigma_\mathrm{g}}}
\newcommand{\pmo}[2]{_{-#1}^{+#2}}

\title[Kinematics and dynamics of the SMC]{On the dynamics of the Small Magellanic Cloud through \\ high-resolution ASKAP \hi\ observations}

\author[Di Teodoro et al.]{Enrico\ M.\ Di Teodoro$^{1}$\thanks{E-mail: enrico.diteodoro@anu.edu.au},
N.\ M.\ McClure-Griffiths$^{1}$,
K.\ E.\ Jameson$^{1}$,
H.\ D\'{e}nes$^{1,2,3}$,
\newauthor
John\ M.\ Dickey$^{4}$,
S.\ Stanimirovi\'{c}$^{5}$,
L.\ Staveley-Smith$^{6,7}$,
C.\ Anderson$^{2}$,
J.\ D. Bunton$^{2}$,  
\newauthor
A.\ Chippendale$^{2}$, 
K.\ Lee-Waddell$^{2}$,
A.\ MacLeod$^{2}$,
and M.\ A.\ Voronkov$^{2}$ \\
$^{1}$Research School of Astronomy and Astrophysics - The Australian National University, Canberra, ACT, 2611, Australia\\
$^{2}$CSIRO Astronomy \& Space Science, PO Box 76, Epping NSW 1710 Australia\\
$^{3}$ASTRON - Netherlands Institute for Radio Astronomy, 7991 PD, Dwingeloo, The Netherlands\\
$^{4}$School of Natural Sciences, University of Tasmania, Hobart TAS, Australia\\
$^{5}$Department of Astronomy, University of Wisconsin, Madison, WI 53706, USA\\
$^{6}$International Centre for Radio Astronomy Research (ICRAR), University of Western Australia, Crawley, WA 6009, Australia\\
$^{7}$ARC Centre of Excellence for All Sky Astrophysics in 3 Dimensions (ASTRO 3D)
}

\date{Accepted XXX. Received YYY; in original form ZZZ}

\pubyear{2018}

\begin{document}
\label{firstpage}
\pagerange{\pageref{firstpage}--\pageref{lastpage}}
\maketitle

\begin{abstract}
We use new high-resolution \hi\ data from the Australian Square Kilometre Array Pathfinder (ASKAP) to investigate the dynamics of the Small Magellanic Cloud (SMC). 
We model the \hi\ gas component as a rotating disc of non-negligible angular size, moving into the plane of the sky and undergoing nutation/precession motions. We derive a high-resolution ($\sim10$ pc) rotation curve of the SMC out to $R\sim4 \, \kpc$. 
After correcting for asymmetric drift, the circular velocity slowly rises to a maximum value of $V_\mathrm{c}\simeq 55 \, \kms$ at $R\simeq2.8 \, \kpc$ and possibly flattens outwards.
In spite of the SMC undergoing strong gravitational interactions with its neighbours, its \hi\ rotation curve is akin to that of many isolated gas-rich dwarf galaxies. 
We decompose the rotation curve and explore different dynamical models to deal with the unknown three-dimensional shape of the mass components (gas, stars and dark matter). 
We find that, for reasonable mass-to-light ratios, a dominant dark matter halo with mass $M_\mathrm{DM}(R<4 \, \kpc) \simeq 1-1.5 \times 10^9 \, \mo$ is always required to successfully reproduce the observed rotation curve, implying a large baryon fraction of $30\%-40\%$.
We discuss  the impact of our assumptions and the limitations of deriving the SMC kinematics and dynamics from \hi\ observations.
\end{abstract}




\section{Introduction}

The Large and Small Magellanic Clouds (LMC and SMC) are gas-rich dwarf satellites of the Milky Way (MW).
The interactions of the Magellanic Clouds (MCs) with each other and with the MW \citep{Besla+10,Besla+12} have produced the Magellanic Stream \citep[][]{Mathewson+74}, a trail of gas extending more than 180$\de$ across the sky, including the Leading Arm. 
The ongoing LMC-SMC interaction is furthermore clearly shown by the stream of gas and stars linking the two galaxies, commonly referred to as Magellanic Bridge \citep[e.g.,][]{McGee&Newton81}.
These gravitational interactions have had a significant impact on the history of the MCs \citep[e.g.,][]{Putman+98} and both their present day morphology and dynamics are highly complex and heavily disturbed.
Studying the current properties of the MCs is the key to understand their recent evolution and interaction mechanisms with the MW.

As the least massive component of the interacting system, the SMC is the most easily affected by tidal forces and therefore shows the most peculiar features. 
Stars and gas in the SMC appear to behave very differently from a structural and kinematical point of view. 
On the plane of the sky, most of the stars lie along a bar-like structure, extending from the north-east to the south-west direction, but the real three-dimensional shape of the SMC is still very uncertain. 
Old and intermediate-age stellar populations are thought to be distributed in a spheroid or ellipsoid significantly elongated along the line of sight (LOS) with the north-eastern region of the bar being closer to us than the south-western region \citep[e.g.,][]{Hatzidimitriou&Hawking89,Cioni+00,Glatt+08,Subramanian&Subramanian12}. 
Younger stellar populations seem instead to be configured in a flatter disc-like structure possibly following the LOS orientation of the older stars \citep[e.g.][]{Subramanian&Subramanian15,Jacyszyn-Dobrzeniecka+16,Ripepi+17}. 
It is however still debated whether the stellar components, in particular the young population, have some significant rotational motion or whether the system is primarily supported by velocity dispersion \citep[e.g.,][]{Harris+06, Evans+08,Dobbie+14,vanderMarel+16,Gaia+18}.

Unlike the stars, the gas in the SMC appears to have features in common with a uniformly rotating disc.
Even the very first single-dish radio-observations of atomic hydrogen (\hi) in the SMC \citep{Kerr+54,Hindman+63} revealed that the gas has a significant velocity gradient along the stellar bar, pointing to possible circular motions, and attempts to derive the rotation curve of the SMC were soon made \citep{Hindman67}.
More recently, interferometric data were obtained with the Australia Telescope Compact Array (ATCA) \citep{Staveley-Smith+97}.
This higher resolution dataset ($\sim100''$) revealed the complexity of the distribution and kinematics of the \hi\ gas in the SMC, showing several hundred expanding shells, arcs and filaments underlying the large scale rotation field \citep{Stanimirovic+99}.
Despite that, the ATCA data provided the best rotation curve for the SMC to date: \citet{Stanimirovic+04} showed that the SMC rotation curve linearly rises to a maximum circular velocity $V_\mathrm{c}\simeq60\, \kms$ at a radius $R\simeq3 \, \kpc$ and proposed a dynamical model of the galaxy where stars and gas alone account for the observed rotation without the necessity of dark matter (DM). 
In a following paper, \citet{Bekki+09} revised this dynamical model and concluded that a dominant DM halo is instead needed to explain the rotation curve from \citet{Stanimirovic+04} work.

In this contribution, we update the SMC gas kinematics and dynamics using new high-resolution \hi\ observations carried out with the Australian Square Kilometre Array Pathfinder \citep[ASKAP,][]{DeBoer+09}. 
Using a generic model of a disc of arbitrary angular size and correcting for proper motion and asymmetric drift, we derive the rotation curve of the SMC with the highest linear resolution ever achieved for any extragalactic system ($\sim 10$ pc).
We fit dynamical mass models to the observed rotation curve, testing the impact of assuming different density distributions for star, gas and DM components, and we estimate the mass contribution of the DM halo within 4 kpc from the SMC centre.

The remainder of this paper is structured as follows. Section \ref{sec:data} describes the new ASKAP data and how we derive the kinematic maps that are used in the dynamical analysis of the SMC. 
Section \ref{sec:theory} introduces the mathematical formalism and the set of equations needed to describe the velocity field of a rotating disc of non-negligible angular size onto the sky. 
We derive the rotation curve of the SMC in Section \ref{sec:kinematics} and we decompose it into the contribution of the different mass components in Section \ref{sec:massdec}. 
We compare our findings with previous studies of the gas kinematics in the SMC in Section \ref{sec:comp}.
Approximations and caveats for our analysis and results are discussed in Section \ref{sec:uncert}. 
We finally summarize and conclude in Section \ref{sec:conc}.

\section{Data}
\label{sec:data}

\subsection{Observations and data reduction}

The SMC was observed in the \hi\ 21-cm emission line with ASKAP as part of the Commissioning and Early Science observations. 
Data were obtained with 16 12-metre antennas with baselines between 22 m and 2.3 km during three consecutive nights (3-5 November 2017) for a total of approximately 36 hours of integration time. 
Each antenna has 36 electronically formed beams arranged on a hexagonal grid on the sky, optimised for uniform sensitivity in a configuration called {\em Closepack36}, which gives a field of view of about $20 \deg^2$.
Beams were formed with maximum signal-to-noise ratio weights determined via the method described in \citet{McConnell+16}.
The total bandwidth of the observations is 240 MHz divided into 12960 channels of 18.5 kHz. 
Because we are only interested in a narrow band around the \hi\ line, we immediately extract a sub-band from 1410 to 1430 MHz, which we used for the subsequent processing.

The data are calibrated using the ASKAPsoft Pipeline v0.19.5\footnote{The official release of ASKAPsoft can be found \href{https://doi.org/10.4225/08/5a0a447510141}{here}. We refer to \href{https://www.atnf.csiro.au/computing/software/askapsoft/sdp/docs/current/pipelines/index.html}{this page} for a detailed documentation (including the default parameters) on the ASKAP Processing Pipeline.} (Whiting et al., in prep.), applied to each day of observation individually. 
The ASKAPsoft pipeline has two main calibration steps: (i) bandpass and absolute flux calibration and (ii) complex gain calibration, which relies on self-calibration techniques.  
Observations with each beam of the primary calibrator PKS B1934-638 are used for bandpass and the primary flux scale.  
The per-beam bandpass solution measured on PKS B1934-638 is applied to the raw data.  
To accurately calibrate the complex gains, we mask out the channels containing bright \hi\ emission from the SMC and the Milky Way and form a continuum image of each beam from the remaining channels. 
For this masking we simply use the default $4\sigma_\mathrm{rms}$ flagging threshold, being $\sigma_\mathrm{rms}$ the root mean square (rms).
We iteratively use the continuum images to solve for self-calibration solutions to the complex antenna-based gains.
The self-calibration module of ASKAPSoft has three main steps: 
(i) the \textsc{Selavy} source-finding algorithm is run with an $8\sigma_\mathrm{rms}$ threshold on the continuum image; 
(ii) the identified source components are used as input for the calibration to create a new gain solution; 
(iii) the latest gain table is applied to the data and a new continuum image is produced.
We repeat this self-calibration loop twice before creating the final continuum image and the final gain solutions, which are eventually applied to the unmasked, bandpass and amplitude calibrated \hi\ visibility data. 
The self-calibration is only applied to the phases, while amplitudes are calibrated exclusively on PKS B1934-638. 

The calibrated data are used to image the \hi\ emission line within the \textsc{Miriad} data reduction package \citep{Sault+95}.
We treat independent beams as individual pointings and image them with traditional linear mosaicking techniques and jointly deconvolve.
During imaging, we use a Briggs' visibility robustness parameter of 0.8, which guarantees a good balance between natural and uniform weighting \citep{Briggs95}.
The image is cleaned using the maximum entropy deconvolution algorithm implemented in \textsc{Mosmem} \citep{Sault+96}, which is particularly appropriate for extended emission, unlike traditional Steer-Dewdney-Ito \citep*[SDI,][]{Steer+84} clean.
We allow \textsc{Mosmem} to run for 100 iterations or until the residuals reached the expected noise level of 1 mJy/beam. For most channels the clean converges and no clean boxes are used.
The restored ASKAP datacube was converted to a Local Standard of Rest (LSR) spectral frame of reference and combined in Fourier space with the single-dish Parkes data from the HI4PI survey \citep{HI4PI}.
We note that the HI4PI single-dish dataset is superior both in coverage and sensitivity to the older Parkes data used in previous kinematic studies of the SMC \citep[e.g.,][]{Stanimirovic+99,Stanimirovic+04}.

The final SMC datacube (ASKAP+Parkes) has a pixel size of 7$''$ and a spectral channel width of 3.9 $\kms$, with a restored beam size of 35$''$ $\times$ 27$''$ Full Width at Half Maximum (FWHM). 
The rms noise in brightness temperature per spectral channel is $\sigma_\mathrm{rms} = 0.7\,\mathrm{K}$.

\begin{figure*}
   \label{fig:maps}
	\includegraphics[width=\textwidth]{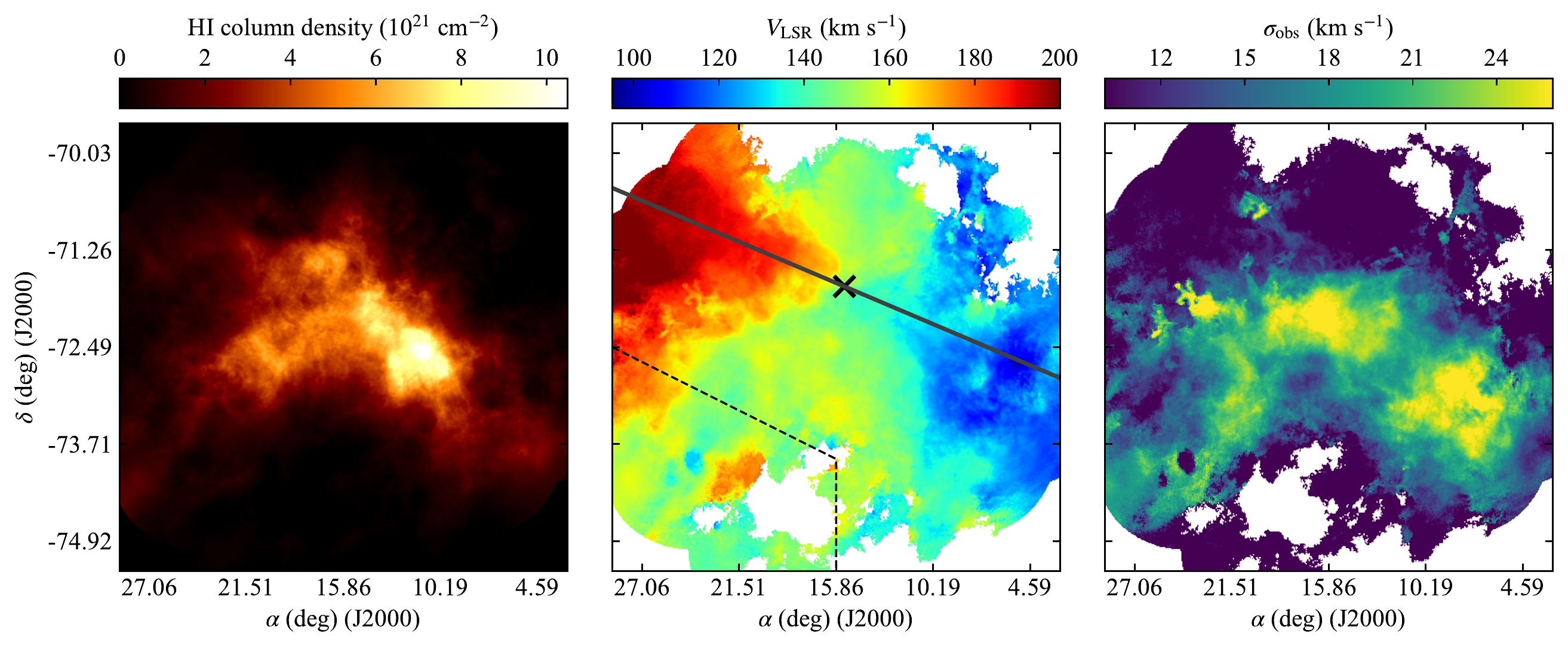}
    \caption{Maps extracted from the \hi\ ASKAP datacube of the SMC. From the left to the right, the column density map, the velocity field and the observed velocity dispersion field. Maps have been derived as described in Section \ref{sec:data}. On the velocity field, we show the best-fit kinematic centre (black cross) and position angle of the line of nodes (dark grey line) derived with the MCMC sampling (Section \ref{sec:globalfit}). The dashed line delimits the region toward the Magellanic Bridge masked out during the entire fitting procedure.}
\end{figure*}

\subsection{Kinematic maps}

We use the ASKAP+Parkes datacube to derive the \hi\ column density map, the velocity and the velocity dispersion fields of the SMC to be used in the kinematic analysis.
Regions of genuine emission are identified through the following procedure. 
We first convolve the datacube with a circular Gaussian kernel with FWHM of 70$''$, i.e.\ twice the beam major axis.
A mask is then obtained using a flood fill algorithm, which starts from pixels with flux higher than 10$\sigma_\mathrm{S}$, where $\sigma_\mathrm{S}$ is the rms of the smoothed datacube, and floods regions down to 4$\sigma_\mathrm{S}$. 
We visually check the mask and apply it to the original datacube to extract the moment maps. 
The $0^\mathrm{th}$ moment $M_0$ is used to obtain the \hi\ column density map, i.e.\ $N_\hi [{\rm cm^{-2}}] = 1.82 \times 10^{18}$ $M_0[{\rm K \, \kms}]$ \citep{Roberts75}, under the assumption that the gas is optically thin.
The $1^\mathrm{st}$ moment (i.e.\ the intensity-weighted mean velocity along the LOS) and the $2^\mathrm{nd}$ moment are taken as velocity field and observed velocity dispersion field, respectively. 

The derived maps are displayed in \autoref{fig:maps}. 
The velocity field (central panel) shows a clear velocity gradient, with the LSR velocities going from about 90 $\kms$ in the south-west region to about 210 $\kms$ in the north-east region.
This strong gradient and the quite regular iso-velocity contours perpendicularly to the major axis suggest that the large-scale motions of the gas are dominated by rotation.
However, the velocity field looks disturbed, especially in the south-east region connecting to the Magellanic Bridge and along the galaxy minor axis, suggestive of non-circular motions possibly related to tidal interactions with the LMC and/or to gas outflowing from the SMC (McClure-Griffiths et al., in press).
To avoid the areas with disturbed kinematics that might hamper our attempts of deriving the underlying regular rotation, in our following kinematic analysis we mask out the south-east region enclosed by the black dashed line in \autoref{fig:maps} (the SMC \vir{wing}) and we apply appropriate weighting functions to down-weigh regions close to the minor axis (see Section \ref{sec:tiltedring}).
The observed velocity dispersion (right panel) varies between 10 $\kms$ and 25 $\kms$, with the highest values towards the Magellanic Bridge and in correspondence of regions of star-formation and supershells \citep{Stanimirovic+99}.
These values include (i) the broadening due to thermal and turbulent motions, i.e.\ the \emph{intrinsic} velocity dispersion of the gas, (ii) the broadening due to geometrical effects when projecting onto the plane of the sky (e.g., in the case of a thick gas layer) and (iii) the instrumental broadening due to the finite spatial and spectral resolutions of a telescope.
We finally stress that the maps shown in \autoref{fig:maps} do not include any correction for geometrical projection effects or proper motions of the SMC: those are directly taken into account during the modelling as described in the next Section.

\section{Theoretical framework}
\label{sec:theory}
Most kinematic studies of external galaxies assume that the observed object  is at large distance, such that its angular size is small and a flat geometry over the area of the galaxy can be assumed \citep[e.g.,][]{Begeman89,Swaters99,deBlok+08}.
Although this is an appropriate assumption for most galaxies, the usual equations used to model the velocity field in distant systems can be unsuitable to properly describe the observed kinematics of both the LMC and the SMC, which respectively subtend about 20\de\ and 6\de\ on the sky.
The geometric formalism to describe the velocity field of a rotating disc of arbitrary angular size was first introduced in the seminal papers by \citet{vdM&Cioni01} and \citet{vdM+02}. 
Here we summarize their main equations and we refer to these previous works for a comprehensive description of the geometry and the derivation of equations.

We consider a generic rotating disc, which is undergoing nutation/precession motions about its symmetry axis and whose centre of mass (CM) has a non-zero transverse motion into the plane of the sky.
The line-of-sight velocity of such a system can be written as \citep[][equation 24]{vdM+02}:

\begin{equation}
\label{eq:vlos1}
\begin{aligned}
V_\mathrm{los}(\rho, \Phi) \,=\, &V_\mathrm{sys}\cos\rho + V_\mathrm{t}\sin\rho\cos(\Phi-\Theta_\mathrm{t}) \\
                   &+ D_0(\didt)\sin\rho\sin(\Phi-\Theta) \\
                   &+ f \vrot (R) \sin i \cos (\Phi-\Theta)
\end{aligned}
\end{equation}

\noindent where the ingredients of the equation are as follows: 

\begin{itemize}
\item $\rho$ and $\Phi$ are angular coordinates that identify the position on the celestial sphere of a point in a frame of reference centered on the CM: $\rho$ is the angular distance from the CM and $\Phi$ is the position angle with respect to the CM, measured anticlockwise from the North direction. 
Spherical trigonometry allows us to derive the angular coordinates from the equatorial coordinates (right ascension $\alpha$ and declination $\delta$) of any point for a fixed position of the CM \citep[][equations 1-3]{vdM&Cioni01}:

\begin{equation}
\label{eq:coords}
\begin{aligned}
\cos\rho &= \cos\delta\cos\delta_0\cos(\alpha-\alpha_0)+\sin\delta\sin\delta_0 \\
\sin\Phi &= \cos\delta\sin(\alpha-\alpha_0)/\sin\rho
\end{aligned}
\end{equation}

\noindent where $(\alpha,\delta)$ and $(\alpha_0,\delta_0)$ are the coordinates of the point and the CM, respectively.\\

\item \vsys, $V_\mathrm{t}$ and $\Theta_\mathrm{t}$ describe the motion of the CM in a frame of reference where the Sun is at rest: \vsys\ is the component along the LOS (positive when receding), often referred to as \emph{systemic velocity}, $V_\mathrm{t}$ is the component perpendicular to the LOS, usually called \emph{transverse velocity}, and $\Theta_\mathrm{t}$ is the angle indicating the direction of the transverse motion on the celestial sphere, measured anticlockwise from North. For the SMC, the transverse velocity causes corrections ranging between $\pm25 \, \kms$ across the velocity field (see Section \ref{sec:globalfit}). \\

\item $D_0$ is the distance of the CM from the Sun. \\

\item $i$ and $\Theta$ describe how the disc plane is viewed from our observing position: $i$ is the inclination angle of the galaxy rotation axis with respect to the LOS ($i=0$ for face-on view) and $\Phi$ is the position angle of the line of nodes (LON), which identifies the intersection of the plane of the galaxy disc with the plane of the sky. In this work, we measure $\Theta$ as the angle between the North direction and the receding part of the LON, taken counterclockwise. \\

\item \didt\ is the time derivative of the inclination angle describing precession and nutation motions. We note that, unlike the inclination angle, variations of the position angle $\Theta$ with time do not affect the observed LOS velocity.\\

\item $f$ is a geometrical factor, defined as:

\begin{equation}
f = \frac{\cos i \cos \rho - \sin i \sin\rho \sin(\Phi-\Theta)}{(\cos^2 i \cos^2 (\Phi-\Theta) +\sin^2 (\Phi-\Theta))^{1/2}} \hspace{10pt}.
\end{equation}
\linebreak

\item \vrot$(R)$ is the rotational velocity in the disc plane at cylindrical radius $R=D_0\sin\rho/f $.\\

\end{itemize}

In practice, \eqref{eq:vlos1} tells us the expected value of the velocity field at each position $(\rho,\Phi)$ for a rotating and precessing disc that is moving across the sky. 
In particular, the first two terms represent the LOS component of the CM velocity vector, the third term is the LOS component due to precession and nutation of the disc and the last term is the LOS component of the rotation. 

Equation \ref{eq:vlos1} can be rewritten to take advantage of the knowledge on the measured galaxy proper motions towards the North $\mu_\mathrm{N} \equiv d\delta /dt$ and towards the West $\mu_\mathrm{W} \equiv -(d\alpha/dt)\cos\delta$ directions. 
Rearranging \eqref{eq:vlos1} leads to \citep[equation 31 in][]{vdM+02}:

\begin{equation}
\label{eq:vlos2}
\begin{aligned}
V_\mathrm{los}(\rho, \Phi) \,=\, &V_\mathrm{sys}\cos\rho + W_\mathrm{ts}\sin\rho\sin(\Phi-\Theta) \\
                   &+ (V_\mathrm{tc}\sin\rho + f \vrot (R) \sin i) \cos (\Phi-\Theta)
\end{aligned}
\end{equation}

\noindent where:

\begin{subequations}
\label{eq:subeq}
\begin{align}
W_\mathrm{ts} &= V_\mathrm{ts} + D_0 (\didt)\\
V_\mathrm{ts} &= V_\mathrm{t} \sin(\Theta_\mathrm{t}-\Theta) = D_0\mu_\mathrm{s}\\
V_\mathrm{tc} &= V_\mathrm{t} \cos(\Theta_\mathrm{t}-\Theta) =  D_0\mu_\mathrm{c} \\
\mu_\mathrm{s} &= -\mu_\mathrm{W}\cos\Theta - \mu_\mathrm{N}\sin\Theta \\
\mu_\mathrm{c} &= -\mu_\mathrm{W}\sin\Theta + \mu_\mathrm{N}\cos\Theta \hspace{10pt}.
\end{align}
\end{subequations}

In the above equations, $V_\mathrm{tc}$ and $V_\mathrm{ts}$ represent the projections of the tangential velocity along and perpendicularly the LON $\Theta$, respectively. 
Similarly, $\mu_\mathrm{c}$ and $\mu_\mathrm{s}$ are the projections of the CM proper-motion vector $(\mu_\mathrm{W},\mu_\mathrm{N})$ along and perpendicular to the LON. 

\section{The \hi\ kinematics of the SMC}
\label{sec:kinematics}
Equations (\ref{eq:vlos1}) and (\ref{eq:vlos2}) are functions of 10 unknown parameters: the centre coordinates  $(\alpha_0, \delta_0)$, the distance $D_0$, the inclination $i$ and LON position angle $\Theta$, the systemic velocity $\vsys$, the transverse velocity $\vtra$ and its direction $\Theta_\mathrm{t}$, the rotation velocity \vrot\ and the precession/nutation term \didt. 
All these quantities but the rotation velocity refer to the disc in its entirety and can in principle be determined by fitting equations (\ref{eq:vlos1}) or (\ref{eq:vlos2}) to the observed velocity field. Conversely, the rotation velocity depends on the radius $R$. 

We therefore divided the problem of fitting the kinematics of the SMC in two steps. 
We first derive the global geometrical and kinematic properties of the SMC disc by comparing the observed velocity field to the modelled \vlos\ via a Monte-Carlo Markov-Chain (MCMC) sampling (Section \ref{sec:globalfit}). 
We then perform a tilted-ring analysis, where we decompose the velocity field in concentric rings at different radii and derive the rotation velocity as a function of $R$ (Section \ref{sec:tiltedring}). 
We finally correct the \hi\ rotation curve for the asymmetric drift and obtain the circular velocity of the SMC (Section \ref{sec:adrift}).

\subsection{Global parameters of the SMC disc}
\label{sec:globalfit}

\begin{figure*}
   \label{fig:corner}
	\includegraphics[width=\textwidth]{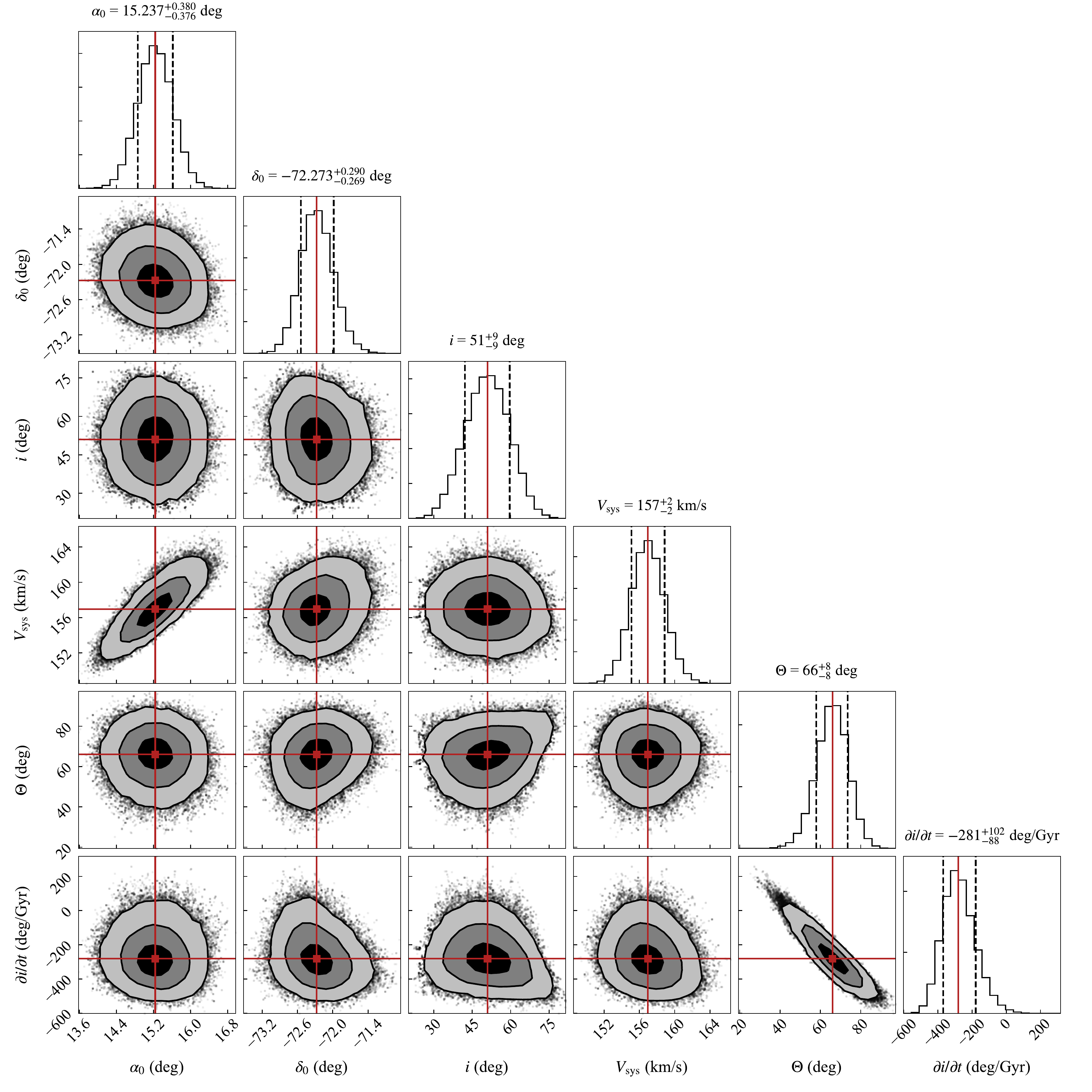}
    \caption{Results of the MCMC sampling. For each pair of parameters, 2D posterior distributions are shown as contours plots. Contours are at 1$\sigma$, 2$\sigma$ and 3$\sigma$ confidence levels. Histograms denote the 1D posterior distributions of each parameter. Full red lines indicate the 50th percentile values, dashed lines are the 15.87th and 84.13th percentiles.}
\end{figure*}

We use \eqref{eq:vlos2} to build simulated velocity fields of the galaxy given a set of parameters and fit them to the observed velocity field. 
In order to reduce the dimensionality of the problem, we take advantage of existing determinations of the SMC distance and proper motions. 
The distance of the SMC from the Sun has been extensively investigated in the literature with several distance ladders, including Cepheids and RR Lyrae \citep[e.g.,][]{Keller&Wood06,Haschke+12}, the tip of the red giant branch \citep[e.g.,][]{Cioni+00,Gorski+16} and eclipsing binaries \citep[][]{North+10,Graczyk+14}. 
In this work we assume a $D_0 = 63 \pm 5$ kpc, representing the median and standard deviation values for a collection of distance estimates in the literature as indicated in the NASA/IPAC Extragalactic Database (NED).
Among the recent proper motion measurements of SMC \citep[e.g.,][]{Piatek+08,Vieira+09,Costa+11}, we decided to use the values $ \mu_\mathrm{W} = -0.772\pm0.063$ mas/yr and $\mu_\mathrm{N} = -1.117\pm0.061$ mas/yr of \cite{Kallivayalil+13}, based on three epochs of Hubble Space Telescope (HST) data spanning 7 years. 
Proper motions are used to calculate the transverse motion parameters $\vtra$, $\Theta_\mathrm{t}$, $V_\mathrm{tc}$ and $V_\mathrm{ts}$ through equations (\ref{eq:subeq}), reducing therefore the degeneracy between $\vrot$, $V_\mathrm{tc}$ and $i$.
We stress that the values of distance and proper motions assumed are much more accurate than what we may achieve with only the \hi\ data.

This first fitting step is meant to derive the global kinematic properties of the SMC disc and does not require a precise knowledge of the variation of the rotation velocity with radius. 
We make use of a simple, empirically-motivated and commonly used parametrization of the rotation curve:

\begin{equation}
\label{eq:vrotpar}
\vrot (R) = \frac{2}{\pi} \, V_\mathrm{f}\,\arctan\left(\frac{R}{R_\mathrm{f}}\right)
\end{equation}

\noindent which rises nearly linearly until some scale radius $R_\mathrm{f}$ and then turns over and flattens to an asymptotic velocity $V_\mathrm{f}$. 
The function in \eqref{eq:vrotpar} can describe the overall rotation pattern of a dwarf galaxy like the SMC with just two parameters, $V_\mathrm{f}$ and $R_\mathrm{f}$.

The comparison between simulated and observed velocity fields is performed via an MCMC sampling, using the python implementation \texttt{emcee} by \citet{Foreman-Mackey+13}. 
Differently from a least square fitting, an MCMC estimates the posterior probability function for unknown parameters and gives a better handling of the correlations and uncertainties.
We recall that the Bayes' theorem states that, given a set of parameters $\mathbf{x}$, the probability $\mathcal{P}(V_\mathrm{mod}(\mathbf{x}) \, | \, V_\mathrm{obs})$ of a model $V_\mathrm{mod}(\mathbf{x})$ given the observable $V_\mathrm{obs}$ is proportional to the product of the \textit{likelihood} $\mathcal{P}(V_\mathrm{obs} \, | \, V_\mathrm{mod}(\mathbf{x}))$ and the \textit{prior} $\mathcal{P}(V_\mathrm{mod}(\mathbf{x}))$. 
An MCMC samples the parameter space $\mathbf{x}$ in a way proportional to the $\mathcal{P}(V_\mathrm{mod}(\mathbf{x}) \, | \, V_\mathrm{obs})$, given the likelihood and the prior functions. 
In our case, $\mathbf{x} = (\alpha_0,\delta_0,\vsys,i,\Theta,\didt,V_\mathrm{f},R_\mathrm{f})$, $V_\mathrm{mod}(\mathbf{x}) \equiv \vlos(\mathbf{x},\rho,\Phi)$ given by \eqref{eq:vlos2} and $V_\mathrm{obs}$ is the observed velocity field (\autoref{fig:maps}).
We assume constant uninformative priors for all parameters and we define the likelihood function as:

\begin{equation}
\label{eq:likelihood}
\log \mathcal{P}(V_\mathrm{obs} \, | \, V_\mathrm{mod}(\mathbf{x})) = -\sum_{j=1}^N \sum_{k=1}^M \frac{(\vlos(\mathbf{x},\rho_i,\Phi_j) - V_\mathrm{obs} (\rho_i,\Phi_j) )^2}{N\times M}
\end{equation}

\noindent where the sums run over all the $\rho$ and $\Phi$ in the data (see \eqref{eq:coords}) and $N\times M$ is the total number of non-blank pixels.
We sample the posterior probability distribution with 500 walkers running 10000 steps for each parameter, including a warm-up phase of 1000 steps. 
We check chains convergence using the integrated autocorrelation time as a diagnostic tool \citep[e.g.,][]{Goodman&Weare10}. 

\autoref{fig:corner} shows 2D marginalized posterior distributions (density plots) and the 1D  joint posterior distributions (histograms on the diagonal) for the MCMC sampling. 
Contours on the 2D distributions show the 1$\sigma$, 2$\sigma$, and 3$\sigma$ confidence levels.
We take the 50th percentile value of the 1D posterior distributions as representative of the central value for each parameter (red-solid lines), while lower and upper errors are taken at the 15.87th and 84.13th percentiles (dashed lines), respectively. 
These values correspond to the mean and the standard deviation for a Gaussian distribution.
Posterior distributions approach a Gaussian function for all parameters, indicating a well-attained sampling. 
The evident correlation between $\Theta$ and $\didt$ is due to the fact that the precession/nutation motions change the position angle of the LON.

The central value and error of each parameter are reported in \autoref{tab:mcmc}. 
Typical 1$\sigma$ uncertainties range between a few percent (e.g. centre, $\vsys$) to about 40\% ($\didt$) of the central values. 
Some parameters estimated with our MCMC sampling appear to slightly differ from those previously derived by \citet{Stanimirovic+04} using \hi\ ATCA observations, although they are consistent within the errors.
In particular, our kinematic centre is offset by $13'$ from theirs, we find a lower systemic velocity (157 vs 160 $\kms$) and larger inclination (51 vs 40 deg) and position angle (66 vs 40 deg). 
The position angle of 40$\de$ used in \citet{Stanimirovic+04} is aligned with the stellar bar, but the \hi\ gas major axis looks offset by more than $20\de$ from that value.
The MCMC sampling returns a precession term of $\mid \didt \mid  = 281 \, \deg \mathrm{Gyr}^{-1}$. This value represents the first attempt to derive the rate of change of the disc inclination angle from \hi\ data and it is a factor two larger than the value of $140 \, \deg \mathrm{Gyr}^{-1}$ found by \citet{Dobbie+14} from the kinematics of red giant stars. 
However, given the large uncertainties associated with this parameter (see \autoref{tab:mcmc} and next paragraph), it is not possible to assert whether stars and gas are truly undergoing different precession motions or not.

Values given in \autoref{tab:mcmc} have been obtained by using proper motions of \cite{Kallivayalil+13}. 
In order to check the dependence of our best kinematic parameters on the assumed proper motions, we repeated the MCMC sampling using the measurements by \citet{Piatek+08}, i.e.\ $\mu_\mathrm{W} = -0.754$ mas yr$^{-1}$ and $\mu_\mathrm{W}= -1.252$ mas yr$^{-1}$. 
We found that the central values of all parameters except $\didt$ are in agreement to within $< 10$\%. 
With proper motions from \citeauthor{Piatek+08}, we obtain a  $\didt = -171\pm120$ deg Gyr$^{-1}$, barely consistent with the one obtained with the values from \citeauthor{Kallivayalil+13}.
This is a further confirmation that the magnitude of precession/nutation motions is not well constrained with the present analysis and should be read as a rough estimate rather than a precise measurement. 
We finally stress that the value of $\didt$ does not affect the derivation of the SMC rotation curve in the next Section, since the precession term in \eqref{eq:vlos1} is null on the kinematic major axis and we exclude regions close to the minor axis.

\begin{table}
\caption{Global properties of the SMC disc, derived through the MCMC sampling described in Section \ref{sec:globalfit}.}
\label{tab:mcmc} 
\centering
\def\arraystretch{1.3}
\begin{tabular}{lc}
\hline\hline\noalign{\vspace{5pt}}
Property & Value \\
\noalign{\smallskip}
\hline\noalign{\vspace{5pt}}
\bf{Assumed:} & \\
Distance $D_0 \,^a$ & $63\pm5$ kpc\\
Proper motion $\mu_\mathrm{W}\,^b$ &  $-0.772\pm0.063$ mas yr$^{-1}$ \\
Proper motion $\mu_\mathrm{N}\,^b$ &  $-1.117\pm0.061$ mas yr$^{-1}$ \\
& \\
\bf{MCMC results:} & \\
RA of kinematic centre $\alpha_0$ (J2000) & $15.237\pmo{0.376}{0.380}$ deg\\
Dec of kinematic centre $\delta_0$ (J2000) & $-72.273\pmo{0.269}{0.290}$ deg\\
Inclination angle $i$ & $51\pm9$ deg \\
Position angle of LON $\Theta\, ^c$ & $66\pm8$ deg \\
Transverse velocity $\vtra ^d$ & $405\pm37$ $\kms$  \\
Transverse direction $\Theta_\mathrm{t} \,^{c,d}$ & $145\pm3$ deg  \\
Systemic velocity $\vsys$ (Heliocentric) & $157 \pm 2$ $\kms$ \\
Systemic velocity $\vsys$ (LSR) & $148 \pm 2$ $\kms$ \\
Precession/Nutation $\didt$ & $-281\pmo{88}{102}$ deg Gyr$^{-1}$ \\
Asymptotic velocity $V_\mathrm{f} \, ^e$ & $56\pm5$ $\kms$ \\
Turnover radius $R_\mathrm{f} \, ^e$ & $1.1\pm0.2$ kpc \\
\noalign{\vspace{2pt}}\hline
\noalign{\vspace{5pt}}
\multicolumn{2}{l}{$^a$Median and standard deviation from the literature (NED).} \\ 
\multicolumn{2}{l}{$^b$From \citet{Kallivayalil+13}.} \\ 
\multicolumn{2}{l}{$^c$Measured anticlockwise from the North direction.} \\ 
\multicolumn{2}{l}{$^d$From proper motions and $\Theta$ through equations (\ref{eq:subeq}).} \\ 
\multicolumn{2}{l}{$^e$For the parametrized rotation curve of \eqref{eq:vrotpar}.} \\ 
\noalign{\vspace{10pt}}
\end{tabular}
\end{table}

\subsection{Tilted-ring model and \hi\ rotation curve}
\label{sec:tiltedring}
We use the parameters derived in the previous Section to fit a tilted-ring model to the observed velocity field and derive the rotation curve of the SMC. 
In a tilted-ring model \citep[e.g.,][]{Rogstad+74,Begeman87}, the galactic disc is decomposed in a number of concentric rings at different radii and each ring orbits about the galaxy centre with a constant rotation velocity. 
Rings are allowed to tilt, i.e.\ to change inclination $i$ and position angle $\Theta$ with radius. 

In this work, we develop a new routine to fit the SMC kinematics with a tilted-ring model. 
In our case the modelled velocity field in each ring is given by \eqref{eq:vlos1}, which is different from all other tilted-ring fitters, like \textsc{Rotcur} \citep[e.g.,][]{vanAlbada+85,Begeman87}, 
The fit to the observed velocity field is performed using a least-square Levenberg-Marquardt solver \citep{Levenberg44,Marquardt63}. 
During the modelling, we fix $\vsys$, $\vtra$, $\Theta_t$ $D_0$ and $\didt$ to the median values previously found (\autoref{tab:mcmc}). 
Following a fairly standard two-step procedure \citep[e.g.][]{Battaglia+06,deBlok+08}, we first perform a tilted-ring analysis allowing the inclination and position angles to vary independently for each ring together with the rotation velocity.
In this step, the values of $i$ and $\Theta$ derived from the MCMC sampling are given as initial guesses for the fit. 
The resulting trends of $i$ and $\Theta$ with radius are then regularized with some parametrized function. 
This is necessary to avoid reflections of non-physical oscillations of the geometrical angles on the derived rotation curve.  
Finally, a second tilted-ring fit is performed, keeping only the rotation velocity free and fixing the inclination and position angles to their best-fit parametrized forms.

\begin{figure}
   \label{fig:tilted}
	\includegraphics[width=0.45\textwidth]{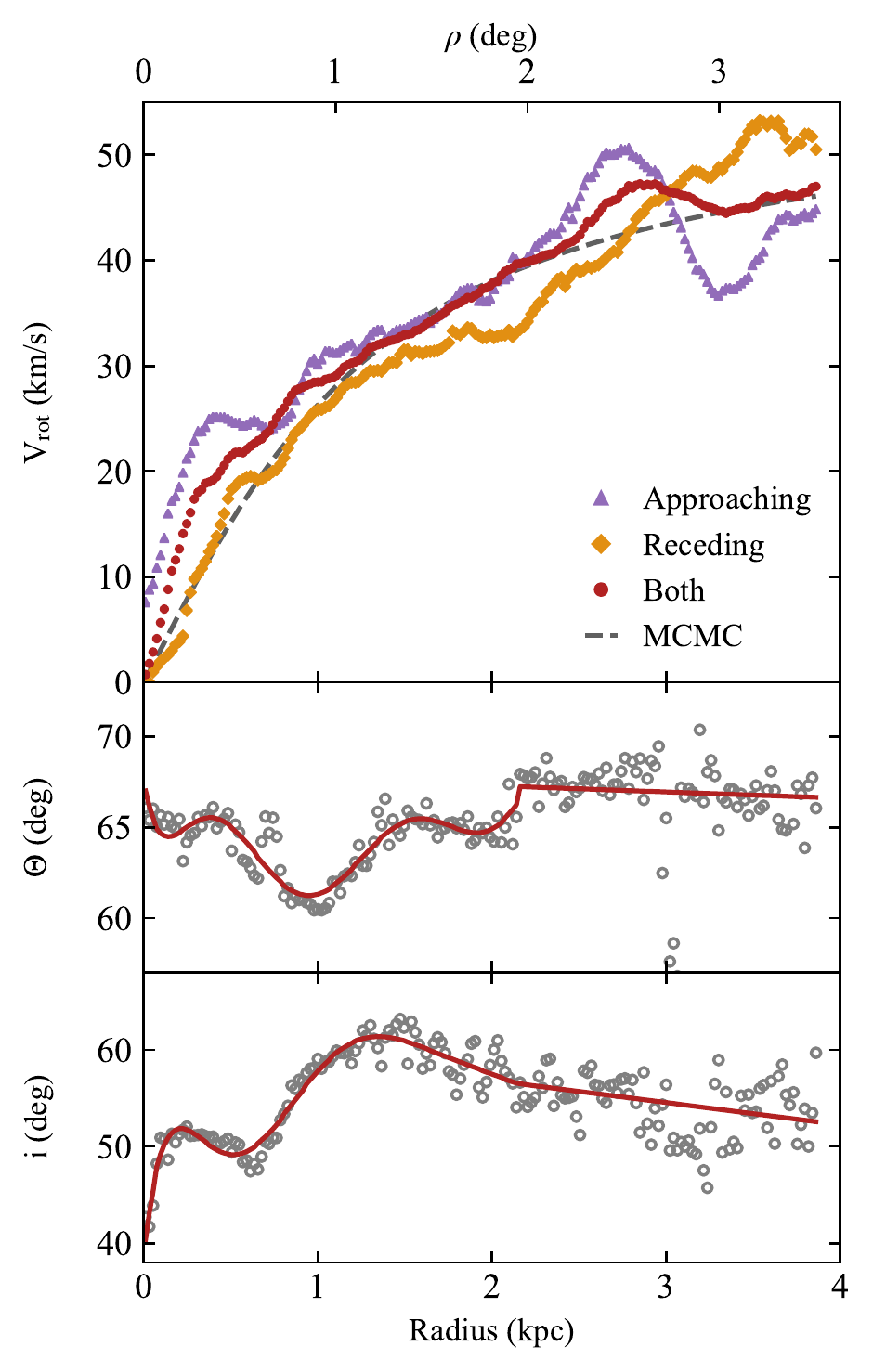}
    \caption{Tilted-ring analysis of SMC. \emph{Top:} rotation curves derived after regularization of the inclination and position angles. Rotation curves of the approaching, receding and both sides of the galaxy are shown with purple triangles, orange diamonds and red circles, respectively. The grey-dashed line denotes the parametrized curve of \eqref{eq:vrotpar} obtained through the MCMC sampling. \emph{Middle:} variation of the position angle of the LON as a function of radius. Grey empty circles represent the first fitting step, the red line is the regularization used to derive the final rotation curve (see Section \ref{sec:tiltedring}). \emph{Bottom:} same as middle panel, but for the inclination angle. The ring width is $35''$, i.e.\ about 10 pc at the distance of the SMC.}
\end{figure}

\begin{figure}
   \label{fig:bestfit}
	\includegraphics[width=0.48\textwidth]{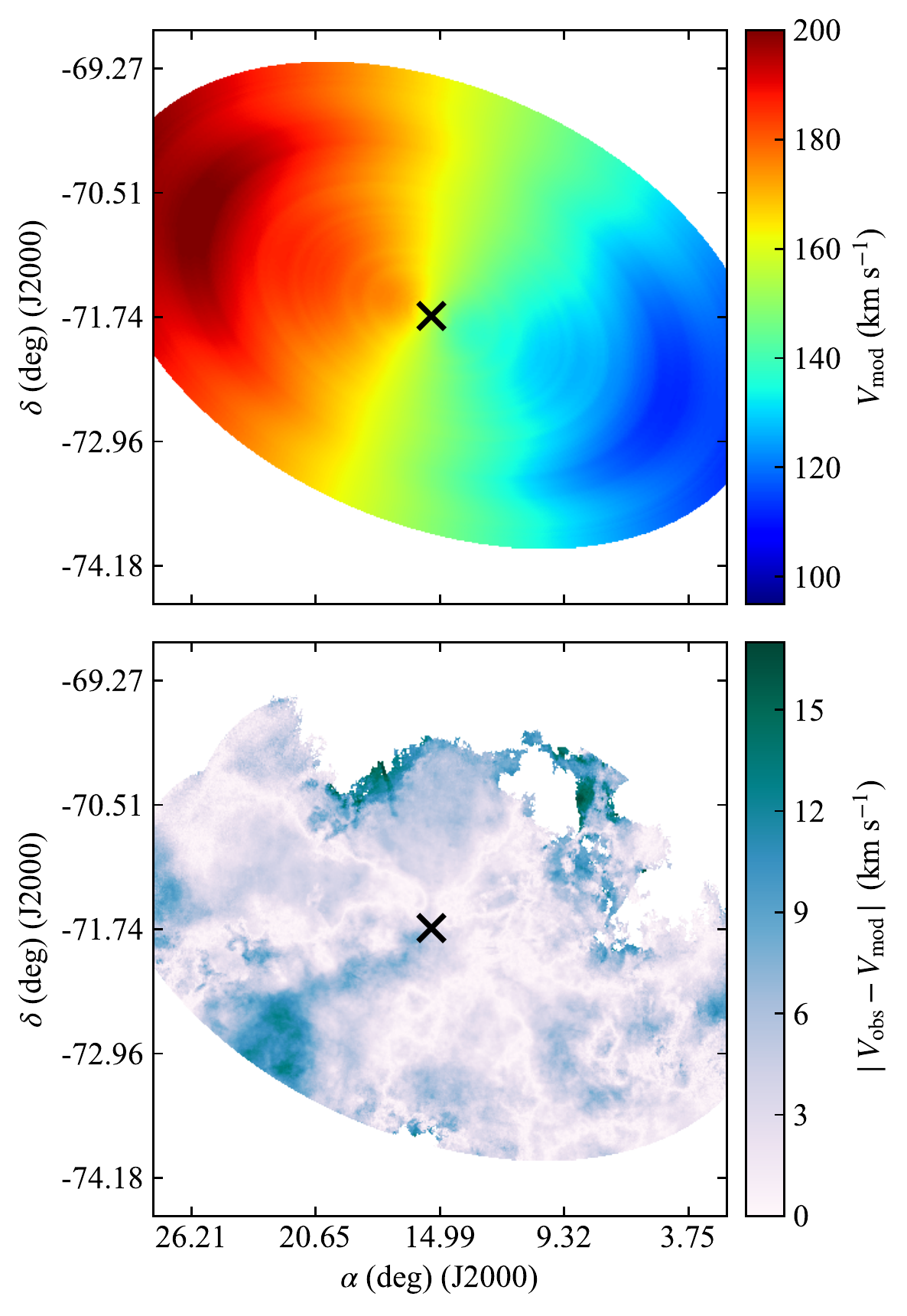}
    \caption{Best-fit tilted ring model of the SMC disc. \emph{Top:} Model velocity field, using the same color scale of the observed velocity field of \autoref{fig:maps}. \emph{Bottom:} Absolute residual map. Residuals are calculated only in the regions where both the observed and modelled velocity field are defined. Black crosses denote the best-fit kinematic centre.}
\end{figure}

During the entire fitting procedure, we set the ring width to 35$''$, i.e.\ equal to the beam size of the observations and corresponding to about 10 pc at the distance of the SMC.
We use a $\cos(\Phi-\Theta)$ weighting function to give more importance to pixels near the kinematic major axis (i.e.\ $\Phi=\Theta$), which is where most of the information on the rotation velocity lies. 
We also exclude regions of the velocity field where $\mid \Phi-\Theta \mid < 20\de$, i.e.\ close to the minor axis. 
After the first fitting step with rotation velocity, position angle and inclination free, we regularize $i$ and $\Theta$ with a 7th-degree polynomial function out to $R=2$ kpc and a straight line for $R>2$ kpc. 
The regularizing functions are chosen to best trace the overall trends of the angles.
The middle and bottom panels of \autoref{fig:tilted} show the solutions of the first fitting step for the position and inclination angles, respectively (grey empty circles) and their regularization (red lines).
The position angle scatters between 60\de\ and 70\de, with a median value very close to the 66\de\ found from the MCMC sampling. 
The inclination angle shows a larger variation in the range $40\de < i < 60\de$, with a median of 55\de\ slightly higher than the central value of the MCMC sampling.

The results of the second fitting step with only the rotation velocity as a free parameter are shown in the top panel of \autoref{fig:tilted}. 
The three rotation curves refer to the approaching half (purple triangles), the receding half (orange diamonds) and both halves (red circles) of the SMC disc. As a reference, we plot also the parametrized rotation curve (equation \ref{eq:vrotpar}) resulting from the MCMC sampling (see \autoref{tab:mcmc}) as a gray dashed line. 
The global rotation curve slowly rises to a maximum velocity of $\sim47$ $\kms$ at $R\sim2.8$ kpc and appears to flatten at larger radii,  although this flattening is not well defined. 
The maximum rotation velocity is slightly larger than the value found in \citet{Stanimirovic+04} ($\sim 40 \, \kms$ and with a lower inclination angle) and older \hi\ studies \citep[$\sim 36 \, \kms$, e.g. ][]{Hindman67,Loiseau81}.
The differences between the rotation curves of the approaching and receding sides reflect the asymmetries of the velocity field. 
The three curves are quite consistent in the region $1 \lesssim R \lesssim 2.5$ kpc, but they show larger discrepancies in the inner and outer parts. 
The latter might be due to the influence of the SMC wing and Magellanic Bridge on the receding side and of the SMC tail and Magellanic Stream on the approaching side.

\autoref{fig:bestfit} shows the velocity field $V_\mathrm{mod}$ of the global best-fit tilted-ring model (top panel) and the absolute residuals between model and observations, i.e.\ $V_\mathrm{res} =  \mid V_\mathrm{obs}-V_\mathrm{mod}\mid$ (bottom panel).
Despite the large asymmetries in the observed velocity field, our completely symmetric tilted-ring model does a fairly good job in reproducing the overall rotation behaviour of the SMC. 
The approaching half seems to be better reproduced than the receding half, which might be more affected by gas in motions to/from the Magellanic Bridge.  
Residuals in regions close to the major axis are of the order of a few $\kms$. 
The largest residuals ($\sim20-25 \, \kms$) are found in regions with $\mid \Phi-\Theta \mid \lesssim 20 \de$, i.e.\ near the minor axis, particularly in the east side. 
These areas are not used during the fit and the large discrepancies from our rotation model may indicate the presence of significant non-circular motions (see discussion in Section \ref{sec:uncert}).
The precession/nutation term $\didt$ also causes a distortion of the velocity field that, according to the third term of \eqref{eq:vlos1}, is more pronounced near the minor axis and at large angular distances from the galaxy centre.
We therefore test the response of the minor axis residuals to different $\didt$ by producing models varying with $-600< \didt < 600$ deg Gyr$^{-1}$.
Consistent with the values found in the MCMC sampling, we find that the more balanced residuals along the minor axis are for $-300 \lesssim \didt \lesssim -100$ deg Gyr$^{-1}$, while more positive/negative values cause larger residuals on one side or the other of the galaxy.

\subsection{Asymmetric drift correction and SMC circular velocity}
\label{sec:adrift}

\begin{figure*}
   \label{fig:adrift}
	\includegraphics[width=\textwidth]{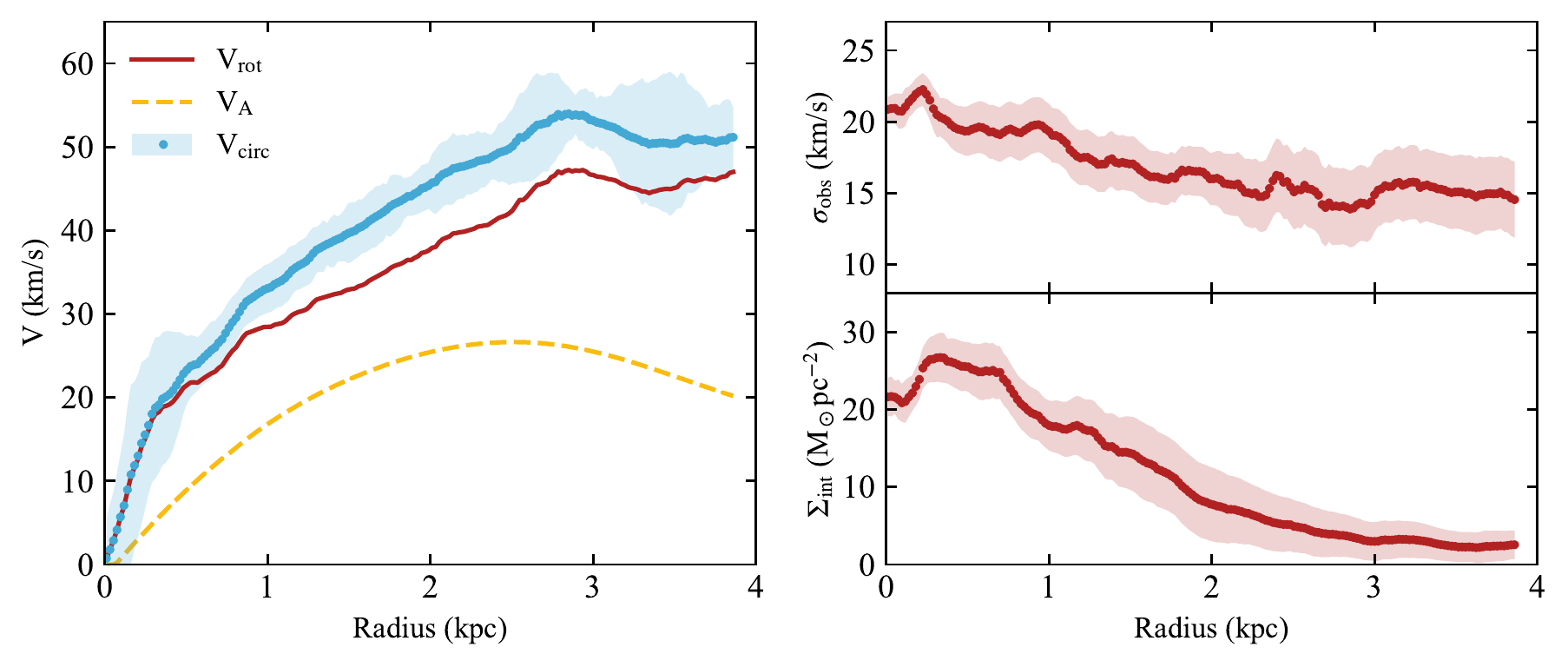}
    \caption{\emph{Left}: Circular velocity of the SMC corrected for the asymmetric drift (cyan dots). Shadowed regions represent the $1-\sigma$ errors, calculated as described in Section \ref{sec:adrift}. The red curve identifies the \hi\ rotation curve derived from both halves of the galaxy (same as \autoref{fig:tilted}), the yellow dashed line the asymmetric drift correction (equation \ref{eq:asym2}). \emph{Right:} velocity dispersion profile (top) and face-on \hi\ mass-density profile derived from the ASKAP data.}
\end{figure*}

The circular velocity $V_\mathrm{c}$, i.e.\ the azimuthal component of the velocity induced in the equatorial plane by an axially symmetric gravitational potential $\phi$, can be written in terms of the average rotation of the gas $\vrot$ in the disc plane plus the so-called \emph{asymmetric drift} correction $V_\mathrm{A}$:

\begin{equation}
\label{eq:vcirc}
V^2_\mathrm{c}(R) = - R\,\frac{\partial \phi (R,z)}{\partial R} \bigg\rvert_{z=0} = V^2_\mathrm{rot}(R) + V^2_\mathrm{A}(R)
\end{equation}

The asymmetric drift correction accounts for the dynamical support due to the turbulent non-ordered motions and can be expressed as a function of the gas velocity dispersion $\vdisp$, the intrinsic surface density $\Sigma_\mathrm{int}$ and the vertical scaleheight $h_\mathrm{z}$ \citep[e.g.][]{Bureau&Carignan02,Oh+11}: 

\begin{equation}
\label{eq:asym}
V^2_\mathrm{A}(R) = -R\sigma^2_\mathrm{g} \frac{\partial \ln (\sigma^2_\mathrm{g} \Sigma_\mathrm{int} h^{-1}_\mathrm{z})}{\partial R} = -R\sigma^2_\mathrm{g} \frac{\partial \ln (\sigma^2_\mathrm{g} \Sigma_\mathrm{obs} \cos i)}{\partial R} \, .
\end{equation}

In the right-hand side of \eqref{eq:asym}, we have made the assumption that the scaleheight is constant, so that we can ignore the radial derivative of $h_\mathrm{z}$, and that the \hi\ disc is razor-thin , which implies $\Sigma_\mathrm{int} = \Sigma_\mathrm{obs}\cos i$, where $\Sigma_\mathrm{obs}$ is the observed surface density.
We also assume that the observed line broadening in each spectrum is a measure of the intrinsic velocity dispersion of the gas.
We discuss the effect of these assumptions in detail in Section \ref{sec:uncert}.

We use the surface density map (\autoref{fig:maps}, left) and the observed velocity dispersion field (\autoref{fig:maps}, right) to extract $\Sigma_\mathrm{obs}(R)$ and $\vdisp(R)$ profiles along the best-fit rings found in Section \ref{sec:tiltedring}. 
We take the median and the standard deviation of velocity dispersion and density in each ring as central values and associated errors. 
Right panels of \autoref{fig:adrift} show the derived surface density (\emph{bottom}) and velocity dispersion (\emph{top}) profiles.
Similarly to many discs of star-forming galaxies, the \hi\ surface density follows an exponential profile with an inner core at $R\lesssim0.8 $ kpc.
The velocity dispersion is quite high at all radii, ranging from about 20 $\kms$ in the central region and dropping to 15 $\kms$ in the outskirts.
The \hi\ velocity dispersion is therefore about half the rotation velocity. 
Assuming that such high values of dispersion are mainly due to turbulence in the SMC disc, then random motions provide a significant support to the system dynamics and the asymmetric drift correction can not be neglected (however, see discussion in Section \ref{sec:uncert}).

The radial derivative in \eqref{eq:asym} is very sensitive to fluctuations in the density and velocity dispersion profiles and it is good practice to regularize both $\vdisp(R)$ and the argument of the logarithmic derivative with analytic functions to avoid meaningless discontinuities in the derived asymmetric drift correction. 
In our case, we regularize the velocity dispersion with a third-degree polynomial $P_3(R)$ and the argument $\sigma^2_\mathrm{g} \Sigma_\mathrm{obs} \cos i$ with a function characterized by an inner core followed by an exponential decline \citep[see e.g.,][]{Iorio+17}:

\begin{equation}
\label{eq:regfunc}
f(R) =  f_0 \frac{R_\mathrm{c}+1}{R_\mathrm{c}+\exp(R/R_\mathrm{d})}
\end{equation}

\noindent where $f_0$ is a normalization factor, $R_\mathrm{c}$ the core radius and $R_\mathrm{d}$ the scale radius for the exponential drop-off.
With these approximations, \eqref{eq:asym} can be conveniently rewritten as

\begin{equation}
\label{eq:asym2}
V^2_\mathrm{A}(R) = \frac{R P^2_3 \exp(R/R_\mathrm{d})}{R_\mathrm{d} (R_\mathrm{c}+\exp(R/R_\mathrm{d}))} \,\, .
\end{equation}

After regularizing the velocity dispersion and $\sigma^2_\mathrm{g} \Sigma_\mathrm{obs} \cos i$, we compute the asymmetric drift correction through \eqref{eq:asym2}. 
We calculate the uncertainties on the final circular velocities by propagating the errors on the rotation velocity and the asymmetric drift based on \eqref{eq:vcirc}. 
The uncertainty on the \hi\ rotation velocity is taken as the statistical error associated with the tilted-ring fit plus half of the difference between the rotation curves of the receding and approaching halves. 
The latter term dominates the error budget in most cases. 
Following \citet{Iorio+17}, we estimate the error on the asymmetric drift correction with a Monte Carlo approach: we produce 1000 realizations of the velocity dispersion $\sigma^i_\mathrm{g}(R)$ and intrinsic density $\Sigma^i_\mathrm{int}(R)$ profiles, where every $i$-th realization is extracted from a normal distribution with mean and standard deviation given by the observed profiles. We then calculate $V^i_\mathrm{A}(R)$ through \eqref{eq:asym2} and take 1.48$\times$MAD($V^i_\mathrm{A}$) as a measure of the asymmetric drift error, where MAD($V^i_\mathrm{A}$) is the median absolute deviation about the median of all realizations.
This procedure allows us to properly account for errors on the derived rotation velocity, velocity dispersion and surface density in the final uncertainties on the derived circular velocity of SMC.

The left panel of \autoref{fig:adrift} shows the \hi\ rotation curve derived for both galaxy halves (red full line), the asymmetric drift correction (yellow dashed line) and the resulting circular velocity through \eqref{eq:vcirc} (cyan dots). 
The shadowed region in \autoref{fig:adrift} represents the uncertainties on the circular velocity formally propagated from the errors on the rotation velocity and asymmetric drift term.
The correction for asymmetric drift causes an increase in the circular velocity of $7-10\, \kms$ between $1\lesssim R \lesssim 2.5$ kpc and of about 5 $\kms$ ($\sim$10\% of the \hi\ rotation velocity) in the outer regions. 
A machine-readable table with our final rotation curve is available as online supplementary material.

\section{Mass decomposition of the SMC}
\label{sec:massdec}
In this Section, we use the asymmetric-drift corrected rotation curve to construct mass models of the SMC. 
The observed circular velocity (equation \ref{eq:vcirc}) can be decomposed into the contribution of the different mass components \citep[e.g.][]{deBlok+08} as

\begin{equation}
\label{eq:vcircdec}
V^2_\mathrm{obs} = V^2_\mathrm{gas} + \Upsilon_*V^2_* + V^2_\mathrm{DM}
\end{equation}

\noindent where $V_\mathrm{gas}$, $V_*$ and $V_\mathrm{DM}$ are the contribution of gas, stars and DM halo to the total rotation velocity. The stellar mass-to-light ratio $\Upsilon_*$ rescales the contribution of the stellar component and is needed because we measure the distribution of light rather than mass. 
In the following sections we describe the density distributions that we use for the various mass components and the assumptions that we make. 
Unless an analytic solution exists, we numerically calculate the circular velocity in the plane $z=0$ induced by any given density distribution integrating \eqref{eq:vcirc}, where the potential $\phi(R,z)$ is computed using the \textsc{Galpynamics} package (Iorio et al., in prep.).

\subsection{Neutral gas distribution}
\label{sec:gasdistr}
We describe the mass surface density distribution of the disc components (both stars and gas) with an exponential multiplied by a polynomial in the radial direction and a squared hyperbolic secant in the vertical direction:

\begin{equation}
\label{eq:thickdisk}
\Sigma_\mathrm{d}(R,z) = \frac{\Sigma_0}{2z_\mathrm{d}} P_\mathrm{n}(R) \exp \left( -\frac{R}{R_\mathrm{d}}\right) \mathrm{sech}^2 \left( \frac{z}{z_\mathrm{d}} \right)
\end{equation}

\noindent where $\Sigma_0$ is the surface density in the centre, $P_\mathrm{n}(R)$ is a $n$-degree polynomial function, $R_\mathrm{d}$ is the disc scalelength and  $h_\mathrm{z}$ is the disc scaleheight. Note that \eqref{eq:thickdisk} implies a constant scaleheight with radius, thus no flaring component is taken into account.
In the case of a razor-thin pure exponential disc, i.e.\ $z_\mathrm{d}=0$ and $P_\mathrm{n} = P_\mathrm{0} = 1$, the circular velocity induced by the disc component is simply given by \citep{Freeman70}:

\begin{equation}
\label{eq:velodisk}
V_\mathrm{d}^2(R) = 4\pi G \Sigma_0 R_\mathrm{d} y^2 \left[ \mathcal{I}_0(y)\mathcal{K}_0(y)-\mathcal{I}_1(y)\mathcal{K}_1(y) \right]
\end{equation}

\noindent where $G$ is the gravitational constant, $y\equiv0.5R/R_\mathrm{d}$ and $\mathcal{I}_\mathrm{n}$ and $\mathcal{K}_\mathrm{n}$ are modified Bessel functions of $n$-th kind. 
For the general case, the circular velocity $V_\mathrm{d}(R)$ needs to be numerically integrated.

For the gas disc, we use a third degree polynomial $P_3(R)$ to describe the observed inner depression in the gas distribution (see \autoref{fig:adrift}, right). 
We correct the \hi\ radial profile by a factor 1.4 to take into account the primordial abundance of helium and metals and fit it for $\Sigma_\mathrm{0,gas}$, $R_\mathrm{d,gas}$ and $P_3(R)$.
We keep the scaleheight $h_\mathrm{z}$ as a free parameter and we build mass model using different scaleheights (see Section \ref{sec:massmod}).

\subsection{Stellar distribution}
\label{sec:stellardistr}
We use two different model distributions for the stellar component: an exponential disc and an exponential prolate spheroid. 
The surface-density distribution of the former is described by \eqref{eq:thickdisk} with $P_n=1$. 
We assume that gas and stellar discs always have the same scaleheight.
The second model is meant to mimic the observed 3D structure of stars in the SMC. 
Several studies have used the period-luminosity relation of variable stars, in particular Classical Cepheids and RR Lyrae, to show that stars in the SMC are distributed in triaxial ellipsoids significantly elongated along the LOS \citep{Deb+15,Jacyszyn-Dobrzeniecka+16,Jacyszyn-Dobrzeniecka+17,Ripepi+17}.
Here we assume the axes ratio 1 : 1.10 : 3.30 quoted in the recent paper of \citet{Muraveva+18}. 
Given the very small ratio of the first two axes, we approximate the triaxial shape with a prolate ellipsoid of axes ratio 1 : 1 : 3.30 and consider the exponential prolate spheroidal density distribution:

\begin{equation}
\label{eq:prolell}
\varrho_\mathrm{star}(m) = \varrho_\mathrm{p}\exp(-m/m_\mathrm{d})
\end{equation}

\noindent where, as usual, $\varrho_\mathrm{p}$ is the central density and $m$ defines the isodensity surfaces $m^2 = R^2 + z^2 / q^2$ with $q\equiv3.30$ in our particular case. The quantity $m_\mathrm{d}$ has the meaning of a scale radius for the variable $m$.

We derive the observed stellar distribution from the mosaicked images at 3.6$\upmu$m of the Surveying the Agents of a Galaxy's Evolution survey of the SMC \citep[SAGE-SMC,][]{Gordon+11}.
SAGE-SMC is a survey of the full SMC system (main body, wing and tail) covering about 30 deg$^2$ in seven bands from 3.6$\upmu$m to 160$\upmu$m with the Infrared Array Camera \citep[IRAC,][]{Fazio+04} and the Multiband Imaging Photometer \citep[MIPS,][]{Rieke+04} on the \emph{Spitzer Space Telescope} \citep{Werner+04}.
At near-infrared (NIR) wavelengths, the light traces the old stellar populations and the effects of dust extinction and star formation (like H\textsc{ii} regions) are quite negligible compared to optical bands. 
Moreover, the mass-to-light ratio $\Upsilon_*$ in the NIR, in particular at 3.6$\upmu$m, is almost constant over a wide-range of galaxy morphologies \citep*[e.g.,][]{Bell&deJong01,Zibetti+09}, reducing the uncertainties related to this unknown parameter.
These characteristics make the 3.6$\upmu$m light optimal to estimate the stellar mass distribution.

We use the high resolution SAGE-SMC images (pixel size = 2$''$) to measure the stellar surface-brightness profile $I_{3.6}(R)$ along the tilted-ring model rings. 
We then calculate the average foreground emission in five regions nearby the SMC and subtract it from the profile.  
We finally convert from surface brightness to mass surface density following \citet{Oh+08}:

\begin{equation}
\label{eq:stellarmass}
\frac{\Sigma_* }{ \mo \pc^{-2}} =  C \Upsilon^{3.6}_*  \cos i \frac{I_{3.6}}{\rm MJy \, sr^{-1}}
\end{equation}

\noindent where $\Upsilon^{3.6}_*$ is the stellar mass-to-light ratio at 3.6$\upmu$m and $C = 696 \, L_\odot$ sr MJy$^{-1}$ pc$^{-2}$. The constant comes from $C = A / Z_{3.6} 10^{[6+0.4(\mathcal{M}^{3.6}_\odot+21.56)]}$, where $A = 2.35 \times 10^{-11}$ sr arcsec$^{-2}$ is the number of steradians in one square arcsec, $Z_{3.6} = 280.9$ Jy is the IRAC zero magnitude flux density at 3.6$\upmu$m \citep{Reach+05} and $\mathcal{M}^{3.6}_\odot = 3.24$ is the absolute magnitude of the Sun at 3.6$\upmu$m \citep[see][for further details]{Oh+08}.
We find that the 3.6$\upmu$m surface-brightness profile is reasonably well described by a pure exponential function.

The mass-to-light ratio $\Upsilon_*$  constitutes the largest uncertainty in the conversion from stellar luminosity to mass \citep[e.g.,][]{Ponomareva+18} and this uncertainty reflects in any rotation curve decomposition. 
Several stellar population synthesis studies have found that the mass-to-light ratio at 3.6$\upmu$m has little dependence on the population age and metallicity, thus on the color \citep{Meidt+14, Norris+14}. 
For example, for a \citet{Kroupa01} initial mass function (IMF), the mass-to-light ratio varies between $0.4 \lesssim \Upsilon^{3.6}_* \lesssim 0.8$ for stellar populations with ages between 3 Gyr and 10 Gyr \citep{Rock+15}, i.e. by a factor of 2. 
Although this variation is small compared to that in the optical \citep[e.g., a factor of 10 within the same age range in $V$ band,][] {Vazdekis+10}, it can still introduce a significant uncertainty in the derived stellar mass. 
In addition, the absolute value of $\Upsilon^{3.6}_*$ is sensitive to the assumed IMF \citep{Rock+15}.
A constant $\Upsilon^{3.6}_*=0.5-0.6$ is often appropriate for star-forming galaxies \citep[e.g.,][]{Meidt+14,McGaugh&Schombert14}. 
A very similar value, $\Upsilon^{3.6}\simeq0.53$, has been also found by \citet*[][]{Eskew+12} for the LMC, based on spatially resolved star formation histories and a \citet{Salpeter55} IMF.
To date, an analogue study on the SMC is still missing. 
In this contribution, we build models for several mass-to-light ratios (see Section \ref{sec:massmod}).

\subsection{Dark matter halos}
We consider two different spherical models for the dark matter distribution.
The first model is a pseudo-isothermal halo, which has a density profile given by:

\begin{equation}
\label{eq:isohalo}
\varrho_\mathrm{iso}(r) = \varrho_0 \left[ 1+ \left( \frac{r}{r_\mathrm{c}} \right)^2 \right]^{-1}
\end{equation}

\noindent where $r$ is the spherical radius, $\varrho_0$ is the central density and $r_\mathrm{c}$ is the core radius of the halo. 
This model is characterized by a central constant-density core and a quadratic decline for radii larger than $r_\mathrm{c}$.
The contribution to the circular velocity due to the density profile in \eqref{eq:isohalo} is

\begin{equation}
\label{eq:isovelo}
V^2_\mathrm{iso}(R) = 4\pi G \varrho_0 r_\mathrm{C}^2 \left[ 1 - \frac{r_\mathrm{c}}{R} \arctan \left(  \frac{R}{r_\mathrm{c}}\right) \right] \,\, .
\end{equation}

The second model is a halo with a Navarro-Frenk-White \citep*[NFW,][]{Navarro+96,Navarro+97} density profile:

\begin{equation}
\label{eq:nfwhalo}
\varrho_\mathrm{NFW}(r) = \varrho_\mathrm{c} \left[  \frac{r}{r_\mathrm{s}} \left(  1+ \frac{r}{r_\mathrm{s}}\right)^2  \right]^{-1}
\end{equation}

\noindent where $r_\mathrm{s}$ is a scale radius of the halo and $\varrho_\mathrm{c} = \varrho_\mathrm{crit}\delta_\mathrm{c}$, being $\varrho_\mathrm{crit}$ the critical density of the Universe and $\delta_\mathrm{c}$ a dimensionless characteristic contrast density.
The NFW profile arises from N-body simulations of structure formation in a Cold Dark Matter (CDM) cosmology and is characterized by a central density cusp ($\varrho_\mathrm{NFW}\sim r^{-1}$ for $r\ll r_\mathrm{s}$). 
As for the pseudo-isothermal case, the rotation velocity in the plane of the disc induced by a NFW halo can be written in analytic form as:

\begin{equation}
\label{eq:nfwvelo}
V^2_\mathrm{NFW} (R) =  V^2_{200} \, \frac{\ln(1+cx)-cx/(1+cx)}{x[\ln(1+c)-c/(1+c)]} 
\end{equation}

\noindent where $x=R/R_{200}$, $V_{200}$ is the rotation velocity at $R_{200}$, the radius inside which the average density is 200 times the critical density of the Universe, roughly corresponding to the virial radius, and $c=R_{200}/R_\mathrm{s}$ is the halo concentration. The parameters $c$ and $V_{200}$ can be then related to density profile parameters $\varrho_\mathrm{c}$ and $r_\mathrm{s}$ \citep[see][]{Navarro+96}.

\begin{figure*}
   \label{fig:massdec_disk}
	\includegraphics[width=\textwidth]{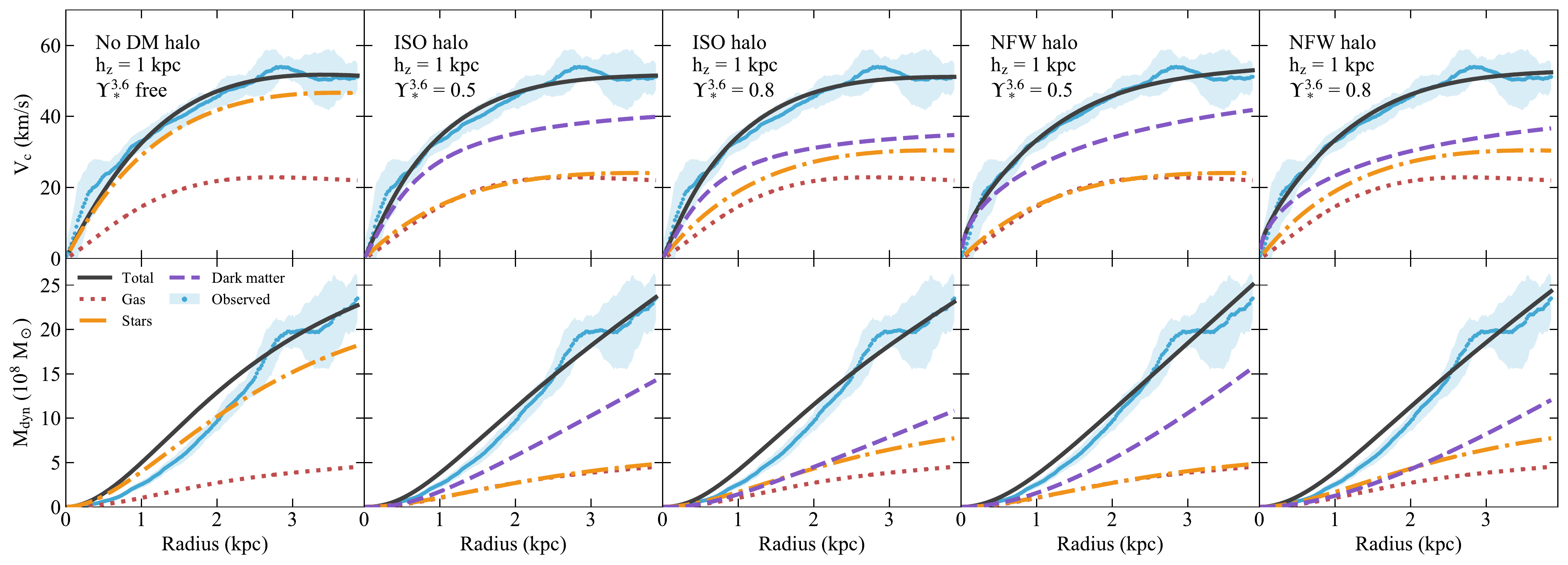}
    \caption{Mass models for the SMC with a thick disc stellar and gas components ($h_\mathrm{z}=1$ kpc). Models with different DM halos (no DM, pseudo-isothermal and NFW) and 3.6$\upmu$m mass-to-light ratios are represented. Top row shows the rotation curve fits for each model, bottom row is the correspondent total mass within a radius $R$. The contributions of stars, gas and dark matter are shown as yellow dotted-dashed, red dotted and purple dashed line, respectively. The total model is the black thick line. Cyan dots represent the observed quantities, with the shadowed areas being the $1-\sigma$ errors. Mass models with no dark matter (first column) are highly unlikely as they lead to unreasonable $\Upsilon^{3.6}_*$. }
\end{figure*}

\subsection{Mass models}
\label{sec:massmod}
We use the observed rotation curve and the derived stellar and gas circular velocities to least-square fit the parameters of the DM halo through \eqref{eq:vcircdec}.
Given the uncertainties on the actual 3D shape of the SMC, we explore a range of interesting cases by varying the properties of both the baryonic and non-baryonic matter.
In particular, we vary the thickness of the disc components in the range $0 \leq h_\mathrm{z} \leq 1$ kpc, the mass-to-light ratio between $0.2 \leq \Upsilon^{3.6}_* \leq 1$, chosen to take into account the wide range of values predicted by stellar population synthesis models \citep[e.g.][]{Rock+15}, and we use either a disc or a prolate spheroid for the stellar component. 
We build models with and without DM: in the first case, we assume either a pseudo-isothermal (cored) or a NFW (cusped) halo and fit their properties to the observed rotation curve, in the second case we let the stellar component supply all the mass needed to reproduce the rotation curve by leaving free the mass-to-light ratio $\Upsilon^{3.6}_*$.

\begin{figure*}
   \label{fig:massdec_prol}
	\includegraphics[width=\textwidth]{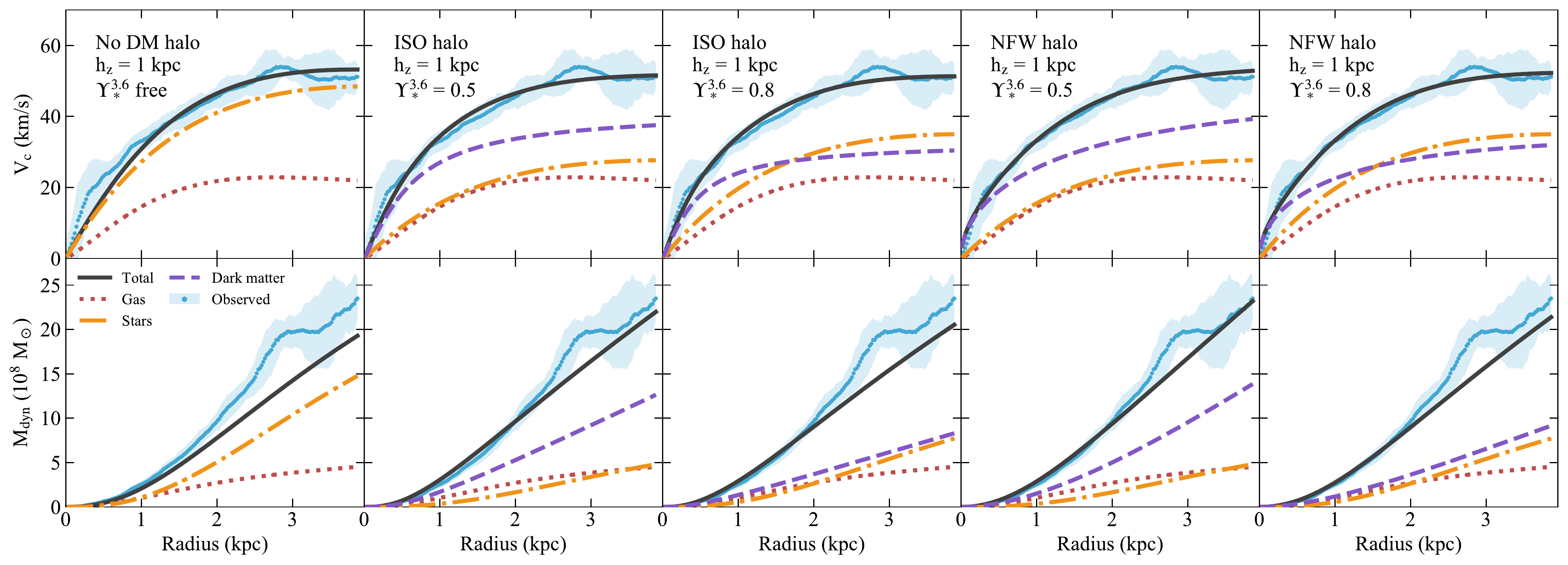}
    \caption{Same as \autoref{fig:massdec_disk}, but for mass models with a thick gas disc ($h_\mathrm{z}=1$ kpc) and a prolate stellar spheroid with axis ratio $q = 3.30$ (see equation \ref{eq:prolell}). Our best model is with a NFW halo and a stellar mass-to-light ratio $\Upsilon^{3.6}_* = 0.5$ (fourth column from the left).}
\end{figure*}

\autoref{fig:massdec_disk} and \autoref{fig:massdec_prol} show mass models with a stellar disc and a stellar prolate ellipsoid, respectively, and a thick disc component $h_\mathrm{z} = 1$ kpc. 
Models with a thin disc ($h_\mathrm{z}\sim0$ kpc) are not shown as they provide a poorer fit to the data.
In both figures, the first row represents the rotation curve decomposition, the second row denotes the corresponding cumulative masses for stars (dotted-dashed yellow line), gas (red dotted) and DM (dashed purple), 
calculated by integrating the best-fit density profiles.
The observed rotation curve and enclosed mass for a spherical distribution $M_\mathrm{obs}(<R) = RV^2_\mathrm{obs}(R)/G$ are shown as blue dots, with shadowed regions representing the errors. 
Black solid lines are the total modelled rotation curve and cumulative mass. 
Starting from the left, we plot models with no DM, pseudo-isothermal and NFW DM halos with fixed $\Upsilon^{3.6}_*=0.5$ and $\Upsilon^{3.6}_*=0.8$.
Most models provide good fits to the observed circular velocity. 
However, models with no DM (first column) lead to very high mass-to-light ratios ($\Upsilon^{3.6}_*\gtrsim 1.5$) and stellar masses ($M_* > 10^9 \, \mo$), which make them unlikely. 
A DM matter halo is needed for all models with sensible $\Upsilon^{3.6}_*$. 
In particular, for $\Upsilon^{3.6}_*<0.7$, the DM component always dominates the overall mass budget at all radii.
The total needed DM mass within 4 kpc is of the order of $1-1.5\times10^9 \, \mo$ depending on the assumed stellar mass-to-light ratio. 
Although NFW halos provide a slightly better fit to the data than pseudo-isothermal halos in terms of $\chi^2$, nothing conclusive can be said about the shape of the inner DM density profile as either halos fail in reproducing the inner rising part of the rotation curve. 
We finally note that the dynamics of the system is not strongly affected by the assumed shape of the stellar component, as either a thick disc (\autoref{fig:massdec_disk}) or a prolate ellipsoid (\autoref{fig:massdec_prol}) lead to similar mass decompositions and to rotation curves that differ by just a few $\kms$, being equal the total enclosed stellar mass.

The best mass model (i.e.\ lowest $\chi^2$) is achieved with the ellipsoidal stellar distribution, a thick gas disc distribution ($h_\mathrm{z} = 1$ kpc), a NFW DM halo and $\Upsilon^{3.6}_*=0.5$ (fourth column in \autoref{fig:massdec_prol}).
The best-fit DM halo has a central density $\varrho_\mathrm{c} = 5.2 \times 10^6 \, \mo \, {\rm kpc^{-3}}$ and a scale radius $r_\mathrm{s} = 5.7$ kpc.
This model nicely reproduces the observed rotation curve and well traces the cumulative mass except for the bump at around 3 kpc. 
A mass-to-light ratio $\Upsilon^{3.6}_*=0.5$ is consistent with stellar population models \citep{Meidt+14, McGaugh&Schombert14} and estimates in the LMC \citep{Eskew+12}.
This implies a stellar mass $M_* \sim 4.8 \times 10^8 \, \mo$ within $R\simeq4$ kpc, slightly higher but comparable with the total estimated stellar mass of the SMC \citep[$\simeq3-3.5\times10^8 \, \mo$, e.g.,][]{Harris+04,Skibba+12}. 
The total mass of neutral gas (\hi\ + He + metals) within the same radius is $M_\mathrm{gas} \sim 4.7 \times 10^8 \, \mo$, of which $M_\hi \sim 3.4 \times 10^8 \, \mo$ is atomic hydrogen \citep[see also][]{Stanimirovic+99}.
A dominant dark matter halo with mass $M_\mathrm{DM} \sim 1.4\times10^9 \, \mo$ is therefore needed to justify a total inferred dynamical mass of $M_\mathrm{tot} \simeq 2.4\times10^9 \, \mo$ (bottom row of \autoref{fig:massdec_prol}). 
The dynamical mass found in this work is consistent with the one quoted in \citet{Stanimirovic+04} but smaller than the value of $2.7-5.1 \times 10^9 \, \mo$ independently estimated by \citet{Harris+06} based on the velocity dispersion of stars.
In our best-fit model, the baryon fraction of the SMC is therefore $f_\mathrm{bar}\sim40\%$, a large value compared to other dwarf galaxies in the local Universe \citep[e.g.,][]{Oh+15}. 
The strong concentration of baryons may indicate that the SMC accreted some gas and stars during the current interaction with the LMC or during a past major merger event \citep{Bekki&Chiba08}. 
Alternatively, the SMC may have recently lost a significant amount of its original DM halo.

\subsection{Mass models in Modified Newtonian dynamics}

Modified Newtonian dynamics \citep[MOND,][for a review]{Milgrom83,Sanders&McGaugh02} is often seen as an alternative to dark matter to explain the discrepancy between theoretical baryon-only and observed rotation curves of galaxies \citep[e.g.,][]{deBlok&McGaugh98,Swaters+10,Randriamampandry&Carignan14}.
In this Section, we decompose the SMC rotation curve using the MOND paradigm and investigate whether MOND can reproduce the observed kinematics of the dwarf galaxy without the need of a DM halo.

Within the MOND framework, the equivalent of \eqref{eq:vcircdec} for the circular velocity can be written as \citep[e.g.,][]{Gentile08}:

\begin{equation}
\label{eq:vcircmond}
V^2_\mathrm{MOND} = V^2_\mathrm{bar}+ V^2_\mathrm{bar}\left(\frac{\sqrt{1+4a_0r/V^2_\mathrm{bar}}-1}{2} \right)
\end{equation}

\noindent where $V^2_\mathrm{bar} = V^2_\mathrm{gas} + \Upsilon_*V^2_*$
is the Newtonian contribution to the circular velocity of baryonic matter (stars + gas) and $a_0$ is the critical acceleration below which the MOND regime dominates and the Newtonian gravity is no longer valid. 
The second term of \eqref{eq:vcircmond}, which is entirely set by the baryonic matter and vanishes for $a_0\rightarrow0$, replaces the DM halo in Newtonian dynamics.
We stress that \eqref{eq:vcircmond} is only valid for an interpolation function $\upmu(x) = x(1+x)^{-1}$ with $x\equiv g/a_0$ and $g=V^2/r$, as proposed by \citet{Famaey&Binney05}.
In this work, we assume the value $a_0 = 1.35\times10^{-8} \, \mathrm{cm \, s^{-2}} = 4166 \, \mathrm{km^2 \, s^{-2} \, kpc^{-1}} $ found by \citet{Famaey+07} for the above interpolation function.

\begin{figure}
   \label{fig:massdec_mond}
	\includegraphics[width=0.49\textwidth]{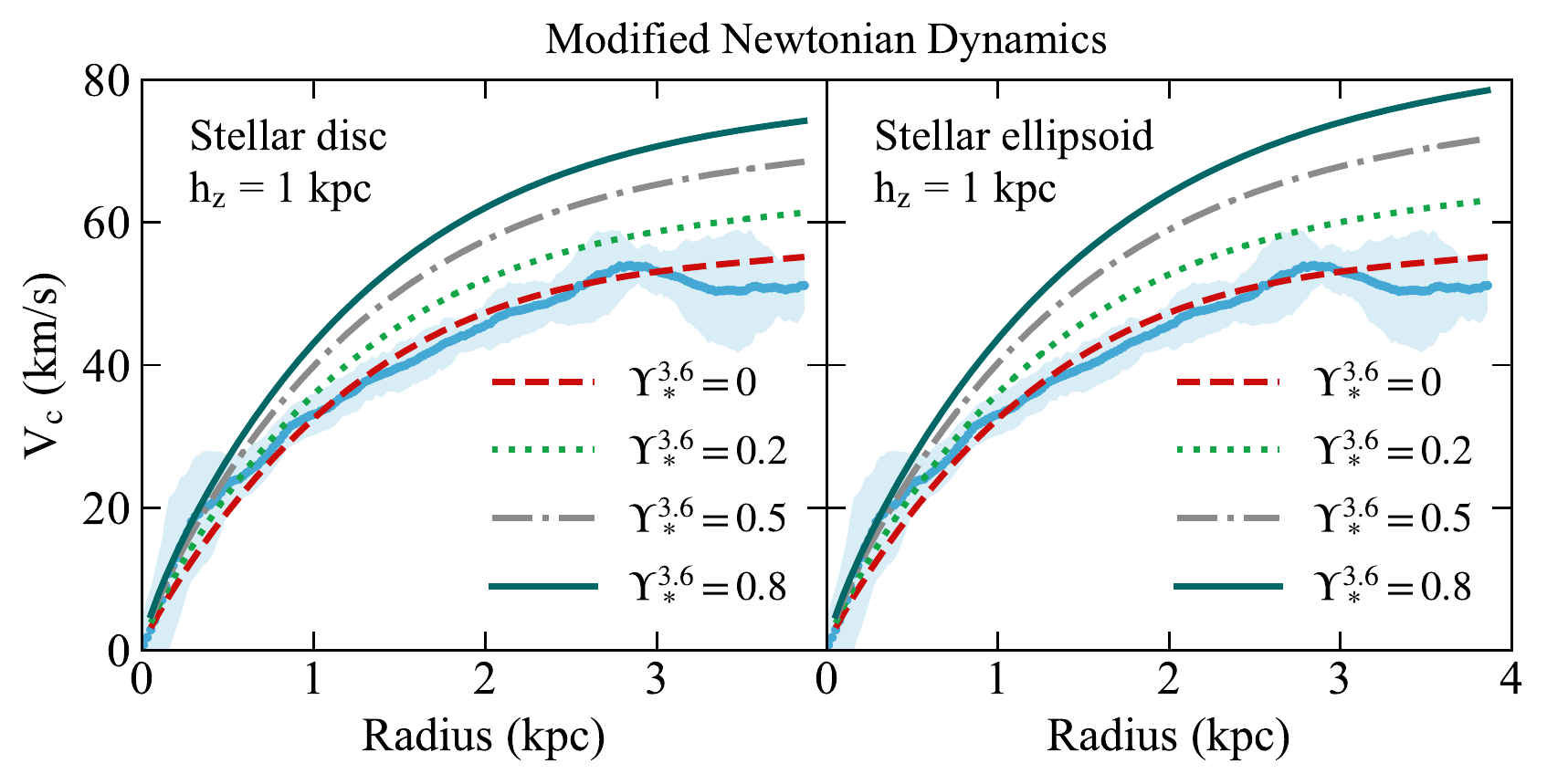}
    \caption{Decomposition of the SMC rotation curve in a MOND framework, using a stellar disc (left) or a stellar prolate ellipsoid (right). 
    The curves denote the MOND predictions for different mass-to-light ratios $\Upsilon^{3.6}_*$, as indicated in the legend. MOND overpredicts the circular velocity of the SMC for $\Upsilon^{3.6}_*>0$.}
\end{figure}

We least-square fit \eqref{eq:vcircmond} to the observed circular velocity curve of the SMC. 
The velocity contributions of gas ($V_\mathrm{gas}$) and stars ($V_*$) are calculated as in the Newtonian case using the mass profiles described in Sections \ref{sec:gasdistr} and \ref{sec:stellardistr}, respectively. 
Since $a_0$ is an universal constant in MOND, we perform the fit with only the mass-to-light ratio $\Upsilon^{3.6}_*$ as a free parameter.
Independently from the disc scaleheight and for either a stellar disc or a stellar ellipsoid, the best fit is achieved for a $\Upsilon^{3.6}_*=0$, i.e.\ nullifying the contribution of the stars to the rotation curve. 
This solution is undoubtedly unphysical and implies that, in a MOND framework, the mass of gas alone is enough to justify the rotation velocity. 
Assuming a non-null $\Upsilon^{3.6}_*$ leads to overestimate the circular velocity at all radii. 
This is clearly shown in \autoref{fig:massdec_mond}, where we plot MOND rotation curves for various $\Upsilon^{3.6}_*$, for both a stellar disc (left panel) and a prolate ellipsoid (right panel) and with a scaleheight $h_\mathrm{z} = 1 \, \kpc$. 
We conclude that the dynamics of the SMC can not be explained in MOND as the galaxy is too rich in baryonic matter, causing MOND to systematically overpredict the circular speed curve for any non-null stellar mass-to-light ratios.

\section{Comparison with other studies}
\label{sec:comp}
The \hi\ kinematics and dynamics of the SMC have been previously investigated in the papers of \citet{Stanimirovic+04} and \citet{Bekki+09}. 
Both those works made use of the same \hi\ dataset obtained from ATCA observations \citep{Staveley-Smith+97} combined with single-dish data from the 64-m Parkes telescope \citep{Stanimirovic+99}.
These data have a smaller sky coverage (20 vs 45 deg$^2$) and velocity coverage (130 vs 510 $\kms$),  a coarser spatial resolution (98$''$ vs 35$''$) and less sensitivity (1.3 vs 0.7 K) than our ASKAP data, but a finer spectral resolution (1.65 vs 3.90 $\kms$).

\citet{Stanimirovic+04} use a proper-motion-corrected velocity field to derive the rotation curve of the SMC with a classic tilted-ring model, including an asymmetric drift correction.
We note that they decided to use only the Parkes \hi\ data as the ATCA+Parkes data were too complex. 
Compared to their approach, we start from updated proper motions and distance (\autoref{tab:mcmc}), we use spherical geometry to take into account the velocity field distortions due to the angular size of the SMC (Section \ref{sec:theory}) and we follow a two-step fitting procedure to separately estimate the global kinematic/geometrical parameters (Section \ref{sec:globalfit}) and the rotation curve (Section \ref{sec:tiltedring}).
\citeauthor{Stanimirovic+04} find geometrical parameters that differ from ours, in particular the kinematic centre, position and inclination angles, which results in slightly different shapes of the observed rotation curves. 
Their maximum observed velocity, i.e.\ before the correction for asymmetric drift, is however similar to the one derived in this work. 
After the asymmetric drift correction, the circular velocity from \citeauthor{Stanimirovic+04} reaches a maximum of $60 \, \kms$, a value larger than ours ($\sim 55 \, \kms$) but consistent within the respective errors.
They decompose the rotation curve with a two-component (gas + star discs) mass model and concluded that no DM halo is needed to explain the observed velocities. 
Their model however implies an excessively large stellar mass $M_* = 1.8\times10^9 \, \mo$ within $R=3.5$ kpc, which conflicts with recent estimates \citep{Skibba+12}.

\citet{Bekki+09} reanalyse the rotation curve from \citet{Stanimirovic+04} to investigate the possible contribution of a DM halo within the central 3 kpc. 
They use $V$-band images to trace the stellar component, they adopt a smaller total stellar luminosity than \citet{Stanimirovic+04} and model the observed rotation curve with a thick disc stellar and gas components and a DM halo following either a NFW or a \citet{Burkert95} density profile.
We note that the usage of $V$-band images to derive the stellar contribution to the total rotation curve introduces a significant uncertainty due to the large variations of the mass-to-light ratio in the optical.
They nonetheless end up with DM halo masses of the order of $1-1.5 \times 10^9 \, \mo$ for reasonable $\Upsilon^V_*$ within 3 kpc, in good agreement with our best-fit value. 
Different from our analysis, they point out that a Burkert DM profile with a large core reproduces the observed rotation curve better than a cusped NFW halo. 
This is likely due to the fact that our rotation curve rises more steeply than that in \citet{Stanimirovic+04} in the inner 0.5 kpc and therefore DM profiles with a central cusp are favoured over profiles with a central core. 

Stellar velocities for the SMC have been widely studied in the past years using several facilities \citep[e.g.,][]{Harris+06, Dobbie+14}. 
In particular, latest data releases from the Gaia satellite are providing improved measurements for proper motions of stars in the Magellanic system \citep{vanderMarel+16, Gaia+18}. 
Gaia results confirm that stars and gas in the SMC do not share common kinematics.
While the gas appears to be settled in a disc with a fairly well-defined rotation curve, the stars do not show a strong and coherent velocity gradient across the galaxy and several kinematic systems with very different dynamical histories likely co-exist depending on the stellar population.
Although proper motions suggest some sense of rotation (certainly less than \hi), attempts to derive a stellar rotation curve are also hindered by the complex and uncertain geometry \citep{Gaia+18}.
Gaia together with previous optical surveys will allow us to better determine how the various stellar systems trace the effects of interaction between the MCs as well as past episodes of enhanced star formation.
Given the complexity of the stellar components, the \hi\ gas is probably the best tracer of the gravitational potential of the SMC, being sufficiently widespread and dynamically relaxed.
Finally, we note that the global proper motions of the SMC determined with Gaia are quite similar to and statistically consistent with the values from \citet{Kallivayalil+13} values used in our analysis. 
As we discussed in Section \ref{sec:globalfit}, slightly different values of  $\mu_\mathrm{N}$ and $\mu_\mathrm{W}$ do not strongly affect our derived kinematics, as the errors on the rotation curve are dominated by the asymmetries between the receding and approaching side and by the inclination uncertainties.

\section{Caveats}
\label{sec:uncert}

Our analysis is based on several techniques widely used in the literature to derive the kinematics of external galaxies from emission-line observations. 
In particular, the tilted-ring model (Section \ref{sec:tiltedring}) and the correction for asymmetric drift (Section \ref{sec:adrift}) are commonly applied to both disc galaxies \citep[e.g.,][]{Begeman87,deBlok+08} and dwarf galaxies \citep[e.g.,][]{Swaters99,Oh+15}.
A tilted-ring model relies on the assumption that gas moves in perfectly circular orbits about the galaxy centre. 
Moreover, when the model is applied to a 2D velocity field, the disc is presumed to be infinitely thin.
Although these assumptions are well suited for spiral galaxies, they can be inappropriate for dwarf galaxies like the MCs, where the shallow gravitational potential causes the disc to be quite thick, especially in the outer regions \citep{Roychowdhury+10}, and non-circular motions can be a non-negligible fraction of the rotation velocity. 
In the case of the SMC, the tidal interactions with the LMC and the MW add a further source of uncertainty.
Although accounting for all these effects is virtually impossible, it is useful to discuss these approximations and the impact they might have on the results presented in this paper. 

The thin disc assumption allows us to easily translate between observed and intrinsic properties of a galaxy: (i) the values on the 2D velocity field directly measure the rotation motions in the equatorial plane, (ii) the face-on surface density can be calculated by simply correcting the observed density profile for the inclination angle and (iii) the observed velocity dispersion is a measure of the intrinsic turbulent motions. 
In the presence of a thick gaseous disc, these properties do not hold and the derived kinematics may not reflect the true kinematics of the galaxy.
We stress that 2D methods based on the velocity maps, like the ones used in this work, can not handle a thick disc.
Three-dimensional techniques \citep{Jozsa+07,DiTeodoro&Fraternali15} guarantee a better treatment of the disc scaleheight, but they are computationally expensive and can not be applied to our large ASKAP data.
However, the effects of approximating a thick with a thin disc are known \citep[e.g.,][]{Iorio+17}: (i) the derived rotation velocity at small (large) radii is lower (higher) than the real rotation velocity, (ii) the observed density profile is shallower than the true profile and (iii) the measured velocity dispersion overestimates the real turbulent motions. 
The first point affects the derived shape of the rotation curve, the last two points lower the correction for the asymmetric-drift (equation \ref{eq:asym}).
In summary, a thick disc causes the circular velocity in the outer regions to be overestimated, which results in the overestimation of the total dynamical mass and, consequently, of the mass of the DM halo.
In this context, our estimated DM mass of $M_\mathrm{DM} \simeq 1.4\times10^9 \, \mo$ should be read as an upper limit.

The second big limitation of our modelling is the assumption of pure circular motions. 
Although the SMC velocity field (\autoref{fig:maps}) shows some clear regularity, it is far from being the classical velocity field of a simply rotating galaxy.
The effect of regular non-circular velocity components, like streaming motions for instance, usually shows up as a twist in the iso-velocity contours, more evident on the minor axis. 
The regions close to the minor axis of the SMC velocity field significantly deviate from regular rotation, suggesting the presence of substantial non-circular motions.
One could in theory modify \eqref{eq:vlos1} to take into account additional velocity components, like radial and/or vertical motions, however this would lead to further degeneracies that would be difficult to control.
Moreover, the velocity residuals (\autoref{fig:bestfit}) show no regular pattern, suggesting that non-circular motions in the SMC are likely dominated by the tidal interaction with the LMC/MW and can not be easily described with additional velocity terms. 
We tried to reduce the impact of non-circular motions by applying a weighting function and masking disturbed regions of the velocity field, it is however difficult to quantify their net effect on the derivation of the rotation curve.  
A secondary consequence of chaotic non-circular motions is the broadening of the line profiles, which results in the turbulent velocity dispersion being overestimated. 
This couples with the analogous effect of the disc thickness and leads to an overcorrection for the asymmetric drift and a circular velocity larger than the real one. 
  
Finally, a potentially important source of errors is the inclination angle, which is very uncertain for the SMC. 
The estimate of the true inclination angle from \hi\ data is hampered by the complex morphology of the SMC and its unknown geometry (for example, a thick disc would cause the inclination angle to be underestimated). 
Independent determinations of the inclination for the putative stellar disc are even more problematic, with values ranging from $5\de$ to $70\de$ depending also on the probed stellar population \citep[e.g.,][]{deVaucouleurs55,Subramanian&Subramanian12,Haschke+12}.
In this work, we obtain an inclination $i=(51\pm9)\de$ from the MCMC sampling and an average $ \langle i \rangle = 55\de$ from the tilted-ring model for the \hi\ disc.
These values are quite a bit larger than the $40\de$ previously found by \citet{Stanimirovic+04}. 
Although our \hi\ data seems to discourage very low inclination angles, in our modelling we still have uncertainties of the order of 10$\de$. 
Since the measured rotation velocity scales as $(\sin i)^{-1}$, an inclination 10$\de$ lower would lead to a larger maximum circular velocity of $\sim 64 \, \kms$ and a larger dynamical mass $\sim 3.2 \times 10^9 \, \mo$.
On the other end, an inclination 10$\de$ higher would translate into a smaller maximum velocity of $\sim 50 \, \kms$ and a smaller dynamical mass $\sim 2 \times 10^9 \, \mo$.

\section{Summary and final remarks}
\label{sec:conc}
New early-science ASKAP observations of the \hi\ content in the SMC allowed us to derive its gas kinematics and study its internal dynamics with unprecedented accuracy. 
The spatial resolution of these data is three times better and the sensitivity is a factor of two higher than former \hi\ interferometric observations of the SMC.
Despite the fact that the velocity field of the SMC appears significantly disturbed because of the gravitational interactions with the LMC, it shows some degree of symmetry and the clear velocity gradient from the north-east to the south-west region is suggestive of differential rotation dominating the large-scale motions.
Our new ASKAP data are ideal to investigate this underlying rotational component.
Different from previous studies of gas kinematics but similar to studies of stellar kinematics in the Magellanic Clouds, we used the mathematical formalism to describe the velocity projection onto the celestial sphere of a rotating disk of generic size and fit this model to the observed velocity field of the SMC. 
Our model takes into account the SMC transverse velocity, through the latest proper motions estimates available, and possible precession of the rotation axis of the SMC disc.

We derived the SMC kinematics in two steps. We first used an MCMC sampling to obtain the global properties of the gas disc, i.e.\ its geometry (centre, inclination and position angles) and motions of the centre of mass (systemic velocity and precession term). 
We then fitted a modified tilted-ring model and derived the rotation curve of the SMC out to $R\sim4\, \kpc$.
We used the \hi\ surface density and velocity dispersion radial profiles extracted from our ASKAP data to compute the asymmetric-drift term and correct the observed rotation curve to account for the contribution of turbulent/random motions to the total dynamical support.
We found that the rotation curve of the SMC slowly rises out to a radius $R\simeq2.8 \, \kpc$, where it reaches a maximum velocity of about $55 \, \kms$, and appears to flatten at larger radii.
Interestingly, the shape of the SMC rotation curve is like that of many isolated gas-rich dwarf galaxies in the local Universe. 
Unlike the stellar kinematics, the gas kinematics probably still preserves the information on what the SMC used to be before the interaction with the LMC and the Milky Way.

We decomposed the rotation curve of the SMC into the contribution of the various dynamical components, namely gas, stars and dark matter. 
To overcome our ignorance on the three-dimensional shape of the SMC, we considered models with either a disc or a prolate spheroid for the stellar component, different scale-heights for the disc components and either a cored pseudo-isothermal or a cusped Navarro-Frenk-White dark matter halo.
Additionally, we explored a range of stellar mass-to-light ratios.
We ruled out the possibility that the SMC has no dark matter halo because this would lead to very high mass-to-light ratios and stellar masses that are in disagreement with our current knowledge of the SMC system.
We additionally showed that alternative gravities, in particular MOND, fail in reproducing the observed kinematics of the galaxy.
A dominant dark matter halo with mass $1-1.5 \times 10^9 \, \mo$ is always necessary to account for a total dynamical mass of $2.4 \times 10^9 \, \mo$ and reproduce the observed rotation curve.
We found the best model is comprised of a thick gas disc, a prolate ellipsoidal stellar component with mass-to-light ratio $\Upsilon^{3.6}_*=0.5$ and a Navarro-Frenk-White dark matter halo, although also models with isothermal halos also provide consistently good fit.
Our best model implies a dark matter mass of $1.4 \times 10^9 \, \mo$ and a baryon fraction of $40\%$ within 4 kpc.
This latter value is significantly larger than what has been found in dwarf galaxies in the local Universe, suggesting that the SMC has either accreted baryonic matter or lost part of its dark matter halo during the tidal interactions with the LMC and the MW.

The dynamics of the Magellanic Clouds can provide important insights into the recent evolution of the Magellanic system.
However, we are probably still distant from an accurate and conclusive understanding of the internal dynamics of the SMC. 
Although \hi\ observations are a powerful tracer of the gas motions, the derivation of the intrinsic SMC kinematics is affected and limited by effects that are difficult to deal with, such as the complex and unknown geometry of the system and the significant non-circular motions induced by tidal forces. 
New facilities, such as ASKAP and Gaia, will help us advance our knowledge of the kinematic and structural properties of gas and stars in the Magellanic Clouds and shed light upon the past history of our satellite galaxies.

\section*{Acknowledgements}
E.D.T.\ and N.M.-G.\ acknowledge the support of the Australian Research Council (ARC) through grant DP160100723. N.M.-G.\ acknowledges the support of the ARC through Future Fellowship FT150100024.
This research has made use of NED, which is operated by the Jet Propulsion Laboratory, California Institute of Technology, under contract with the National Aeronautics and Space Administration.
The Australian SKA Pathfinder is part of the Australia Telescope National Facility which is managed by CSIRO. Operation of ASKAP is funded by the Australian Government with support from the National Collaborative Research Infrastructure Strategy. ASKAP uses the resources of the Pawsey Supercomputing Centre. Establishment of ASKAP, the Murchison Radio-astronomy Observatory and the Pawsey Supercomputing Centre are initiatives of the Australian Government, with support from the Government of Western Australia and the Science and Industry Endowment Fund. We acknowledge the Wajarri Yamatji people as the traditional owners of the Observatory site.


\bibliographystyle{mnras}
\bibliography{biblio}

\begin{thebibliography}{}
\makeatletter
\relax
\def\mn@urlcharsother{\let\do\@makeother \do\$\do\&\do\#\do\^\do\_\do\%\do\~}
\def\mn@doi{\begingroup\mn@urlcharsother \@ifnextchar [ {\mn@doi@}
  {\mn@doi@[]}}
\def\mn@doi@[#1]#2{\def\@tempa{#1}\ifx\@tempa\@empty \href
  {http://dx.doi.org/#2} {doi:#2}\else \href {http://dx.doi.org/#2} {#1}\fi
  \endgroup}
\def\mn@eprint#1#2{\mn@eprint@#1:#2::\@nil}
\def\mn@eprint@arXiv#1{\href {http://arxiv.org/abs/#1} {{\tt arXiv:#1}}}
\def\mn@eprint@dblp#1{\href {http://dblp.uni-trier.de/rec/bibtex/#1.xml}
  {dblp:#1}}
\def\mn@eprint@#1:#2:#3:#4\@nil{\def\@tempa {#1}\def\@tempb {#2}\def\@tempc
  {#3}\ifx \@tempc \@empty \let \@tempc \@tempb \let \@tempb \@tempa \fi \ifx
  \@tempb \@empty \def\@tempb {arXiv}\fi \@ifundefined
  {mn@eprint@\@tempb}{\@tempb:\@tempc}{\expandafter \expandafter \csname
  mn@eprint@\@tempb\endcsname \expandafter{\@tempc}}}

\bibitem[\protect\citeauthoryear{{Battaglia}, {Fraternali}, {Oosterloo}  \&
  {Sancisi}}{{Battaglia} et~al.}{2006}]{Battaglia+06}
{Battaglia} G.,  {Fraternali} F.,  {Oosterloo} T.,   {Sancisi} R.,  2006,
  \mn@doi [\aap] {10.1051/0004-6361:20053210}, \href
  {http://adsabs.harvard.edu/abs/2006A%26A...447...49B} {447, 49}

\bibitem[\protect\citeauthoryear{{Begeman}}{{Begeman}}{1987}]{Begeman87}
{Begeman} K.~G.,  1987, PhD thesis, Kapteyn Institute

\bibitem[\protect\citeauthoryear{{Begeman}}{{Begeman}}{1989}]{Begeman89}
{Begeman} K.~G.,  1989, \aap, \href
  {http://adsabs.harvard.edu/abs/1989A%26A...223...47B} {223, 47}

\bibitem[\protect\citeauthoryear{{Bekki} \& {Chiba}}{{Bekki} \&
  {Chiba}}{2008}]{Bekki&Chiba08}
{Bekki} K.,  {Chiba} M.,  2008, \mn@doi [\apjl] {10.1086/589441}, \href
  {http://adsabs.harvard.edu/abs/2008ApJ...679L..89B} {679, L89}

\bibitem[\protect\citeauthoryear{{Bekki} \& {Stanimirovi{\'c}}}{{Bekki} \&
  {Stanimirovi{\'c}}}{2009}]{Bekki+09}
{Bekki} K.,  {Stanimirovi{\'c}} S.,  2009, \mn@doi [\mnras]
  {10.1111/j.1365-2966.2009.14514.x}, \href
  {http://adsabs.harvard.edu/abs/2009MNRAS.395..342B} {395, 342}

\bibitem[\protect\citeauthoryear{{Bell} \& {de Jong}}{{Bell} \& {de
  Jong}}{2001}]{Bell&deJong01}
{Bell} E.~F.,  {de Jong} R.~S.,  2001, \mn@doi [\apj] {10.1086/319728}, \href
  {http://adsabs.harvard.edu/abs/2001ApJ...550..212B} {550, 212}

\bibitem[\protect\citeauthoryear{{Besla}, {Kallivayalil}, {Hernquist}, {van der
  Marel}, {Cox}  \& {Kere{\v s}}}{{Besla} et~al.}{2010}]{Besla+10}
{Besla} G.,  {Kallivayalil} N.,  {Hernquist} L.,  {van der Marel} R.~P.,  {Cox}
  T.~J.,   {Kere{\v s}} D.,  2010, \mn@doi [\apjl]
  {10.1088/2041-8205/721/2/L97}, \href
  {http://adsabs.harvard.edu/abs/2010ApJ...721L..97B} {721, L97}

\bibitem[\protect\citeauthoryear{{Besla}, {Kallivayalil}, {Hernquist}, {van der
  Marel}, {Cox}  \& {Kere{\v s}}}{{Besla} et~al.}{2012}]{Besla+12}
{Besla} G.,  {Kallivayalil} N.,  {Hernquist} L.,  {van der Marel} R.~P.,  {Cox}
  T.~J.,   {Kere{\v s}} D.,  2012, \mn@doi [\mnras]
  {10.1111/j.1365-2966.2012.20466.x}, \href
  {http://adsabs.harvard.edu/abs/2012MNRAS.421.2109B} {421, 2109}

\bibitem[\protect\citeauthoryear{{Briggs}}{{Briggs}}{1995}]{Briggs95}
{Briggs} D.~S.,  1995, PhD thesis, The New Mexico Institute of Mining and
  Technology

\bibitem[\protect\citeauthoryear{{Bureau} \& {Carignan}}{{Bureau} \&
  {Carignan}}{2002}]{Bureau&Carignan02}
{Bureau} M.,  {Carignan} C.,  2002, \mn@doi [\aj] {10.1086/338899}, \href
  {http://adsabs.harvard.edu/abs/2002AJ....123.1316B} {123, 1316}

\bibitem[\protect\citeauthoryear{{Burkert}}{{Burkert}}{1995}]{Burkert95}
{Burkert} A.,  1995, \mn@doi [\apjl] {10.1086/309560}, \href
  {http://adsabs.harvard.edu/abs/1995ApJ...447L..25B} {447, L25}

\bibitem[\protect\citeauthoryear{{Cioni}, {van der Marel}, {Loup}  \&
  {Habing}}{{Cioni} et~al.}{2000}]{Cioni+00}
{Cioni} M.-R.~L.,  {van der Marel} R.~P.,  {Loup} C.,   {Habing} H.~J.,  2000,
  \aap, \href {http://adsabs.harvard.edu/abs/2000A%26A...359..601C} {359, 601}

\bibitem[\protect\citeauthoryear{{Costa}, {M{\'e}ndez}, {Pedreros}, {Moyano},
  {Gallart}  \& {No{\"e}l}}{{Costa} et~al.}{2011}]{Costa+11}
{Costa} E.,  {M{\'e}ndez} R.~A.,  {Pedreros} M.~H.,  {Moyano} M.,  {Gallart}
  C.,   {No{\"e}l} N.,  2011, \mn@doi [\aj] {10.1088/0004-6256/141/4/136},
  \href {http://adsabs.harvard.edu/abs/2011AJ....141..136C} {141, 136}

\bibitem[\protect\citeauthoryear{{DeBoer} et~al.,}{{DeBoer}
  et~al.}{2009}]{DeBoer+09}
{DeBoer} D.~R.,  et~al., 2009, \mn@doi [IEEE Proceedings]
  {10.1109/JPROC.2009.2016516}, \href
  {http://adsabs.harvard.edu/abs/2009IEEEP..97.1507D} {97, 1507}

\bibitem[\protect\citeauthoryear{{Deb}, {Singh}, {Kumar}  \& {Kanbur}}{{Deb}
  et~al.}{2015}]{Deb+15}
{Deb} S.,  {Singh} H.~P.,  {Kumar} S.,   {Kanbur} S.~M.,  2015, \mn@doi
  [\mnras] {10.1093/mnras/stv358}, \href
  {http://adsabs.harvard.edu/abs/2015MNRAS.449.2768D} {449, 2768}

\bibitem[\protect\citeauthoryear{{Di Teodoro} \& {Fraternali}}{{Di Teodoro} \&
  {Fraternali}}{2015}]{DiTeodoro&Fraternali15}
{Di Teodoro} E.~M.,  {Fraternali} F.,  2015, \mn@doi [\mnras]
  {10.1093/mnras/stv1213}, \href
  {http://adsabs.harvard.edu/abs/2015MNRAS.451.3021D} {451, 3021}

\bibitem[\protect\citeauthoryear{{Dobbie}, {Cole}, {Subramaniam}  \&
  {Keller}}{{Dobbie} et~al.}{2014}]{Dobbie+14}
{Dobbie} P.~D.,  {Cole} A.~A.,  {Subramaniam} A.,   {Keller} S.,  2014, \mn@doi
  [\mnras] {10.1093/mnras/stu910}, \href
  {http://adsabs.harvard.edu/abs/2014MNRAS.442.1663D} {442, 1663}

\bibitem[\protect\citeauthoryear{{Eskew}, {Zaritsky}  \& {Meidt}}{{Eskew}
  et~al.}{2012}]{Eskew+12}
{Eskew} M.,  {Zaritsky} D.,   {Meidt} S.,  2012, \mn@doi [\aj]
  {10.1088/0004-6256/143/6/139}, \href
  {http://adsabs.harvard.edu/abs/2012AJ....143..139E} {143, 139}

\bibitem[\protect\citeauthoryear{{Evans} \& {Howarth}}{{Evans} \&
  {Howarth}}{2008}]{Evans+08}
{Evans} C.~J.,  {Howarth} I.~D.,  2008, \mn@doi [\mnras]
  {10.1111/j.1365-2966.2008.13012.x}, \href
  {http://adsabs.harvard.edu/abs/2008MNRAS.386..826E} {386, 826}

\bibitem[\protect\citeauthoryear{{Famaey} \& {Binney}}{{Famaey} \&
  {Binney}}{2005}]{Famaey&Binney05}
{Famaey} B.,  {Binney} J.,  2005, \mn@doi [\mnras]
  {10.1111/j.1365-2966.2005.09474.x}, \href
  {http://adsabs.harvard.edu/abs/2005MNRAS.363..603F} {363, 603}

\bibitem[\protect\citeauthoryear{{Famaey}, {Gentile}, {Bruneton}  \&
  {Zhao}}{{Famaey} et~al.}{2007}]{Famaey+07}
{Famaey} B.,  {Gentile} G.,  {Bruneton} J.-P.,   {Zhao} H.,  2007, \mn@doi
  [\prd] {10.1103/PhysRevD.75.063002}, \href
  {http://adsabs.harvard.edu/abs/2007PhRvD..75f3002F} {75, 063002}

\bibitem[\protect\citeauthoryear{{Fazio} et~al.,}{{Fazio}
  et~al.}{2004}]{Fazio+04}
{Fazio} G.~G.,  et~al., 2004, \mn@doi [\apjs] {10.1086/422843}, \href
  {http://adsabs.harvard.edu/abs/2004ApJS..154...10F} {154, 10}

\bibitem[\protect\citeauthoryear{{Foreman-Mackey}, {Hogg}, {Lang}  \&
  {Goodman}}{{Foreman-Mackey} et~al.}{2013}]{Foreman-Mackey+13}
{Foreman-Mackey} D.,  {Hogg} D.~W.,  {Lang} D.,   {Goodman} J.,  2013, \mn@doi
  [\pasp] {10.1086/670067}, \href
  {http://adsabs.harvard.edu/abs/2013PASP..125..306F} {125, 306}

\bibitem[\protect\citeauthoryear{{Freeman}}{{Freeman}}{1970}]{Freeman70}
{Freeman} K.~C.,  1970, \mn@doi [\apj] {10.1086/150474}, \href
  {http://adsabs.harvard.edu/abs/1970ApJ...160..811F} {160, 811}

\bibitem[\protect\citeauthoryear{{Gaia Collaboration} et~al.,}{{Gaia
  Collaboration} et~al.}{2018}]{Gaia+18}
{Gaia Collaboration} et~al., 2018, \mn@doi [\aap]
  {10.1051/0004-6361/201832698}, \href
  {http://adsabs.harvard.edu/abs/2018A%26A...616A..12G} {616, A12}

\bibitem[\protect\citeauthoryear{{Gentile}}{{Gentile}}{2008}]{Gentile08}
{Gentile} G.,  2008, \mn@doi [\apj] {10.1086/590048}, \href
  {http://adsabs.harvard.edu/abs/2008ApJ...684.1018G} {684, 1018}

\bibitem[\protect\citeauthoryear{{Glatt} et~al.,}{{Glatt}
  et~al.}{2008}]{Glatt+08}
{Glatt} K.,  et~al., 2008, \mn@doi [\aj] {10.1088/0004-6256/136/4/1703}, \href
  {http://adsabs.harvard.edu/abs/2008AJ....136.1703G} {136, 1703}

\bibitem[\protect\citeauthoryear{{Goodman} \& {Weare}}{{Goodman} \&
  {Weare}}{2010}]{Goodman&Weare10}
{Goodman} J.,  {Weare} J.,  2010, \mn@doi [Communications in Applied
  Mathematics and Computational Science, Vol.~5, No.~1, p.~65-80, 2010]
  {10.2140/camcos.2010.5.65}, \href
  {http://adsabs.harvard.edu/abs/2010CAMCS...5...65G} {5, 65}

\bibitem[\protect\citeauthoryear{{Gordon} et~al.,}{{Gordon}
  et~al.}{2011}]{Gordon+11}
{Gordon} K.~D.,  et~al., 2011, \mn@doi [\aj] {10.1088/0004-6256/142/4/102},
  \href {http://adsabs.harvard.edu/abs/2011AJ....142..102G} {142, 102}

\bibitem[\protect\citeauthoryear{{G{\'o}rski} et~al.,}{{G{\'o}rski}
  et~al.}{2016}]{Gorski+16}
{G{\'o}rski} M.,  et~al., 2016, \mn@doi [\aj] {10.3847/0004-6256/151/6/167},
  \href {http://adsabs.harvard.edu/abs/2016AJ....151..167G} {151, 167}

\bibitem[\protect\citeauthoryear{{Graczyk} et~al.,}{{Graczyk}
  et~al.}{2014}]{Graczyk+14}
{Graczyk} D.,  et~al., 2014, \mn@doi [\apj] {10.1088/0004-637X/780/1/59}, \href
  {http://adsabs.harvard.edu/abs/2014ApJ...780...59G} {780, 59}

\bibitem[\protect\citeauthoryear{{HI4PI Collaboration} et~al.,}{{HI4PI
  Collaboration} et~al.}{2016}]{HI4PI}
{HI4PI Collaboration} et~al., 2016, \mn@doi [\aap]
  {10.1051/0004-6361/201629178}, \href
  {http://adsabs.harvard.edu/abs/2016A%26A...594A.116H} {594, A116}

\bibitem[\protect\citeauthoryear{{Harris} \& {Zaritsky}}{{Harris} \&
  {Zaritsky}}{2004}]{Harris+04}
{Harris} J.,  {Zaritsky} D.,  2004, \mn@doi [\aj] {10.1086/381953}, \href
  {http://adsabs.harvard.edu/abs/2004AJ....127.1531H} {127, 1531}

\bibitem[\protect\citeauthoryear{{Harris} \& {Zaritsky}}{{Harris} \&
  {Zaritsky}}{2006}]{Harris+06}
{Harris} J.,  {Zaritsky} D.,  2006, \mn@doi [\aj] {10.1086/500974}, \href
  {http://adsabs.harvard.edu/abs/2006AJ....131.2514H} {131, 2514}

\bibitem[\protect\citeauthoryear{{Haschke}, {Grebel}  \& {Duffau}}{{Haschke}
  et~al.}{2012}]{Haschke+12}
{Haschke} R.,  {Grebel} E.~K.,   {Duffau} S.,  2012, \mn@doi [\aj]
  {10.1088/0004-6256/144/4/107}, \href
  {http://adsabs.harvard.edu/abs/2012AJ....144..107H} {144, 107}

\bibitem[\protect\citeauthoryear{{Hatzidimitriou} \&
  {Hawkins}}{{Hatzidimitriou} \& {Hawkins}}{1989}]{Hatzidimitriou&Hawking89}
{Hatzidimitriou} D.,  {Hawkins} M.~R.~S.,  1989, \mn@doi [\mnras]
  {10.1093/mnras/241.4.667}, \href
  {http://adsabs.harvard.edu/abs/1989MNRAS.241..667H} {241, 667}

\bibitem[\protect\citeauthoryear{{Hindman}}{{Hindman}}{1967}]{Hindman67}
{Hindman} J.~V.,  1967, \mn@doi [Australian Journal of Physics]
  {10.1071/PH670147}, \href {http://adsabs.harvard.edu/abs/1967AuJPh..20..147H}
  {20, 147}

\bibitem[\protect\citeauthoryear{{Hindman}, {Kerr}  \& {McGee}}{{Hindman}
  et~al.}{1963}]{Hindman+63}
{Hindman} J.~V.,  {Kerr} F.~J.,   {McGee} R.~X.,  1963, \mn@doi [Australian
  Journal of Physics] {10.1071/PH630570}, \href
  {http://adsabs.harvard.edu/abs/1963AuJPh..16..570H} {16, 570}

\bibitem[\protect\citeauthoryear{{Iorio}, {Fraternali}, {Nipoti}, {Di Teodoro},
  {Read}  \& {Battaglia}}{{Iorio} et~al.}{2017}]{Iorio+17}
{Iorio} G.,  {Fraternali} F.,  {Nipoti} C.,  {Di Teodoro} E.,  {Read} J.~I.,
  {Battaglia} G.,  2017, \mn@doi [\mnras] {10.1093/mnras/stw3285}, \href
  {http://adsabs.harvard.edu/abs/2017MNRAS.466.4159I} {466, 4159}

\bibitem[\protect\citeauthoryear{{Jacyszyn-Dobrzeniecka}
  et~al.,}{{Jacyszyn-Dobrzeniecka} et~al.}{2016}]{Jacyszyn-Dobrzeniecka+16}
{Jacyszyn-Dobrzeniecka} A.~M.,  et~al., 2016, \actaa, \href
  {http://adsabs.harvard.edu/abs/2016AcA....66..149J} {66, 149}

\bibitem[\protect\citeauthoryear{{Jacyszyn-Dobrzeniecka}
  et~al.,}{{Jacyszyn-Dobrzeniecka} et~al.}{2017}]{Jacyszyn-Dobrzeniecka+17}
{Jacyszyn-Dobrzeniecka} A.~M.,  et~al., 2017, \actaa, \href
  {http://adsabs.harvard.edu/abs/2017AcA....67....1J} {67, 1}

\bibitem[\protect\citeauthoryear{{J{\'o}zsa}, {Kenn}, {Klein}  \&
  {Oosterloo}}{{J{\'o}zsa} et~al.}{2007}]{Jozsa+07}
{J{\'o}zsa} G.~I.~G.,  {Kenn} F.,  {Klein} U.,   {Oosterloo} T.~A.,  2007,
  \mn@doi [\aap] {10.1051/0004-6361:20066164}, \href
  {http://adsabs.harvard.edu/abs/2007A%26A...468..731J} {468, 731}

\bibitem[\protect\citeauthoryear{{Kallivayalil}, {van der Marel}, {Besla},
  {Anderson}  \& {Alcock}}{{Kallivayalil} et~al.}{2013}]{Kallivayalil+13}
{Kallivayalil} N.,  {van der Marel} R.~P.,  {Besla} G.,  {Anderson} J.,
  {Alcock} C.,  2013, \mn@doi [\apj] {10.1088/0004-637X/764/2/161}, \href
  {http://adsabs.harvard.edu/abs/2013ApJ...764..161K} {764, 161}

\bibitem[\protect\citeauthoryear{{Keller} \& {Wood}}{{Keller} \&
  {Wood}}{2006}]{Keller&Wood06}
{Keller} S.~C.,  {Wood} P.~R.,  2006, \mn@doi [\apj] {10.1086/501115}, \href
  {http://adsabs.harvard.edu/abs/2006ApJ...642..834K} {642, 834}

\bibitem[\protect\citeauthoryear{{Kerr}, {Hindman}  \& {Robinson}}{{Kerr}
  et~al.}{1954}]{Kerr+54}
{Kerr} F.~J.,  {Hindman} J.~F.,   {Robinson} B.~J.,  1954, \mn@doi [Australian
  Journal of Physics] {10.1071/PH540297}, \href
  {http://adsabs.harvard.edu/abs/1954AuJPh...7..297K} {7, 297}

\bibitem[\protect\citeauthoryear{{Kroupa}}{{Kroupa}}{2001}]{Kroupa01}
{Kroupa} P.,  2001, \mn@doi [\mnras] {10.1046/j.1365-8711.2001.04022.x}, \href
  {http://adsabs.harvard.edu/abs/2001MNRAS.322..231K} {322, 231}

\bibitem[\protect\citeauthoryear{{Levenberg}}{{Levenberg}}{1944}]{Levenberg44}
{Levenberg} K.,  1944, Quart. Appl. Math., 2, 164

\bibitem[\protect\citeauthoryear{{Loiseau} \& {Bajaja}}{{Loiseau} \&
  {Bajaja}}{1981}]{Loiseau81}
{Loiseau} N.,  {Bajaja} E.,  1981, \rmxaa, \href
  {http://adsabs.harvard.edu/abs/1981RMxAA...6...55L} {6, 55}

\bibitem[\protect\citeauthoryear{{Marquardt}}{{Marquardt}}{1963}]{Marquardt63}
{Marquardt} D.,  1963, SIAM J. Appl. Math., 11, 431

\bibitem[\protect\citeauthoryear{{Mathewson}, {Cleary}  \&
  {Murray}}{{Mathewson} et~al.}{1974}]{Mathewson+74}
{Mathewson} D.~S.,  {Cleary} M.~N.,   {Murray} J.~D.,  1974, \mn@doi [\apj]
  {10.1086/152875}, \href {http://adsabs.harvard.edu/abs/1974ApJ...190..291M}
  {190, 291}

\bibitem[\protect\citeauthoryear{{McConnell} et~al.,}{{McConnell}
  et~al.}{2016}]{McConnell+16}
{McConnell} D.,  et~al., 2016, \mn@doi [\pasa] {10.1017/pasa.2016.37}, \href
  {http://adsabs.harvard.edu/abs/2016PASA...33...42M} {33, e042}

\bibitem[\protect\citeauthoryear{{McGaugh} \& {Schombert}}{{McGaugh} \&
  {Schombert}}{2014}]{McGaugh&Schombert14}
{McGaugh} S.~S.,  {Schombert} J.~M.,  2014, \mn@doi [\aj]
  {10.1088/0004-6256/148/5/77}, \href
  {http://adsabs.harvard.edu/abs/2014AJ....148...77M} {148, 77}

\bibitem[\protect\citeauthoryear{{McGee} \& {Newton}}{{McGee} \&
  {Newton}}{1981}]{McGee&Newton81}
{McGee} R.~X.,  {Newton} L.~M.,  1981, Proceedings of the Astronomical Society
  of Australia, \href {http://adsabs.harvard.edu/abs/1981PASAu...4..189M} {4,
  189}

\bibitem[\protect\citeauthoryear{{Meidt} et~al.,}{{Meidt}
  et~al.}{2014}]{Meidt+14}
{Meidt} S.~E.,  et~al., 2014, \mn@doi [\apj] {10.1088/0004-637X/788/2/144},
  \href {http://adsabs.harvard.edu/abs/2014ApJ...788..144M} {788, 144}

\bibitem[\protect\citeauthoryear{{Milgrom}}{{Milgrom}}{1983}]{Milgrom83}
{Milgrom} M.,  1983, \mn@doi [\apj] {10.1086/161131}, \href
  {http://adsabs.harvard.edu/abs/1983ApJ...270..371M} {270, 371}

\bibitem[\protect\citeauthoryear{{Muraveva} et~al.,}{{Muraveva}
  et~al.}{2018}]{Muraveva+18}
{Muraveva} T.,  et~al., 2018, \mn@doi [\mnras] {10.1093/mnras/stx2514}, \href
  {http://adsabs.harvard.edu/abs/2018MNRAS.473.3131M} {473, 3131}

\bibitem[\protect\citeauthoryear{{Navarro}, {Frenk}  \& {White}}{{Navarro}
  et~al.}{1996}]{Navarro+96}
{Navarro} J.~F.,  {Frenk} C.~S.,   {White} S.~D.~M.,  1996, \mn@doi [\apj]
  {10.1086/177173}, \href {http://adsabs.harvard.edu/abs/1996ApJ...462..563N}
  {462, 563}

\bibitem[\protect\citeauthoryear{{Navarro}, {Frenk}  \& {White}}{{Navarro}
  et~al.}{1997}]{Navarro+97}
{Navarro} J.~F.,  {Frenk} C.~S.,   {White} S.~D.~M.,  1997, \mn@doi [\apj]
  {10.1086/304888}, \href {http://adsabs.harvard.edu/abs/1997ApJ...490..493N}
  {490, 493}

\bibitem[\protect\citeauthoryear{{Norris}, {Meidt}, {Van de Ven}, {Schinnerer},
  {Groves}  \& {Querejeta}}{{Norris} et~al.}{2014}]{Norris+14}
{Norris} M.~A.,  {Meidt} S.,  {Van de Ven} G.,  {Schinnerer} E.,  {Groves} B.,
   {Querejeta} M.,  2014, \mn@doi [\apj] {10.1088/0004-637X/797/1/55}, \href
  {http://adsabs.harvard.edu/abs/2014ApJ...797...55N} {797, 55}

\bibitem[\protect\citeauthoryear{{North}, {Gauderon}, {Barblan}  \&
  {Royer}}{{North} et~al.}{2010}]{North+10}
{North} P.,  {Gauderon} R.,  {Barblan} F.,   {Royer} F.,  2010, \mn@doi [\aap]
  {10.1051/0004-6361/200810284}, \href
  {http://adsabs.harvard.edu/abs/2010A%26A...520A..74N} {520, A74}

\bibitem[\protect\citeauthoryear{{Oh}, {de Blok}, {Walter}, {Brinks}  \&
  {Kennicutt}}{{Oh} et~al.}{2008}]{Oh+08}
{Oh} S.-H.,  {de Blok} W.~J.~G.,  {Walter} F.,  {Brinks} E.,   {Kennicutt} Jr.
  R.~C.,  2008, \mn@doi [\aj] {10.1088/0004-6256/136/6/2761}, \href
  {http://adsabs.harvard.edu/abs/2008AJ....136.2761O} {136, 2761}

\bibitem[\protect\citeauthoryear{{Oh}, {de Blok}, {Brinks}, {Walter}  \&
  {Kennicutt}}{{Oh} et~al.}{2011}]{Oh+11}
{Oh} S.-H.,  {de Blok} W.~J.~G.,  {Brinks} E.,  {Walter} F.,   {Kennicutt} Jr.
  R.~C.,  2011, \mn@doi [\aj] {10.1088/0004-6256/141/6/193}, \href
  {http://adsabs.harvard.edu/abs/2011AJ....141..193O} {141, 193}

\bibitem[\protect\citeauthoryear{{Oh} et~al.,}{{Oh} et~al.}{2015}]{Oh+15}
{Oh} S.-H.,  et~al., 2015, \mn@doi [\aj] {10.1088/0004-6256/149/6/180}, \href
  {http://adsabs.harvard.edu/abs/2015AJ....149..180O} {149, 180}

\bibitem[\protect\citeauthoryear{{Piatek}, {Pryor}  \& {Olszewski}}{{Piatek}
  et~al.}{2008}]{Piatek+08}
{Piatek} S.,  {Pryor} C.,   {Olszewski} E.~W.,  2008, \mn@doi [\aj]
  {10.1088/0004-6256/135/3/1024}, \href
  {http://adsabs.harvard.edu/abs/2008AJ....135.1024P} {135, 1024}

\bibitem[\protect\citeauthoryear{{Ponomareva}, {Verheijen}, {Papastergis},
  {Bosma}  \& {Peletier}}{{Ponomareva} et~al.}{2018}]{Ponomareva+18}
{Ponomareva} A.~A.,  {Verheijen} M.~A.~W.,  {Papastergis} E.,  {Bosma} A.,
  {Peletier} R.~F.,  2018, \mn@doi [\mnras] {10.1093/mnras/stx3066}, \href
  {http://adsabs.harvard.edu/abs/2018MNRAS.474.4366P} {474, 4366}

\bibitem[\protect\citeauthoryear{{Putman} et~al.,}{{Putman}
  et~al.}{1998}]{Putman+98}
{Putman} M.~E.,  et~al., 1998, \mn@doi [\nat] {10.1038/29466}, \href
  {http://adsabs.harvard.edu/abs/1998Natur.394..752P} {394, 752}

\bibitem[\protect\citeauthoryear{{Randriamampandry} \&
  {Carignan}}{{Randriamampandry} \&
  {Carignan}}{2014}]{Randriamampandry&Carignan14}
{Randriamampandry} T.~H.,  {Carignan} C.,  2014, \mn@doi [\mnras]
  {10.1093/mnras/stu100}, \href
  {http://adsabs.harvard.edu/abs/2014MNRAS.439.2132R} {439, 2132}

\bibitem[\protect\citeauthoryear{{Reach} et~al.,}{{Reach}
  et~al.}{2005}]{Reach+05}
{Reach} W.~T.,  et~al., 2005, \mn@doi [\pasp] {10.1086/432670}, \href
  {http://adsabs.harvard.edu/abs/2005PASP..117..978R} {117, 978}

\bibitem[\protect\citeauthoryear{{Rieke} et~al.,}{{Rieke}
  et~al.}{2004}]{Rieke+04}
{Rieke} G.~H.,  et~al., 2004, \mn@doi [\apjs] {10.1086/422717}, \href
  {http://adsabs.harvard.edu/abs/2004ApJS..154...25R} {154, 25}

\bibitem[\protect\citeauthoryear{{Ripepi} et~al.,}{{Ripepi}
  et~al.}{2017}]{Ripepi+17}
{Ripepi} V.,  et~al., 2017, \mn@doi [\mnras] {10.1093/mnras/stx2096}, \href
  {http://adsabs.harvard.edu/abs/2017MNRAS.472..808R} {472, 808}

\bibitem[\protect\citeauthoryear{{Roberts}}{{Roberts}}{1975}]{Roberts75}
{Roberts} M.~S.,  1975, {Radio Observations of Neutral Hydrogen in Galaxies}.
the University of Chicago Press, p.~309

\bibitem[\protect\citeauthoryear{{R{\"o}ck}, {Vazdekis}, {Peletier}, {Knapen}
  \& {Falc{\'o}n-Barroso}}{{R{\"o}ck} et~al.}{2015}]{Rock+15}
{R{\"o}ck} B.,  {Vazdekis} A.,  {Peletier} R.~F.,  {Knapen} J.~H.,
  {Falc{\'o}n-Barroso} J.,  2015, \mn@doi [\mnras] {10.1093/mnras/stv503},
  \href {http://adsabs.harvard.edu/abs/2015MNRAS.449.2853R} {449, 2853}

\bibitem[\protect\citeauthoryear{{Rogstad}, {Lockhart}  \& {Wright}}{{Rogstad}
  et~al.}{1974}]{Rogstad+74}
{Rogstad} D.~H.,  {Lockhart} I.~A.,   {Wright} M.~C.~H.,  1974, \mn@doi [\apj]
  {10.1086/153164}, \href {http://adsabs.harvard.edu/abs/1974ApJ...193..309R}
  {193, 309}

\bibitem[\protect\citeauthoryear{{Roychowdhury}, {Chengalur}, {Begum}  \&
  {Karachentsev}}{{Roychowdhury} et~al.}{2010}]{Roychowdhury+10}
{Roychowdhury} S.,  {Chengalur} J.~N.,  {Begum} A.,   {Karachentsev} I.~D.,
  2010, \mn@doi [\mnras] {10.1111/j.1745-3933.2010.00835.x}, \href
  {http://adsabs.harvard.edu/abs/2010MNRAS.404L..60R} {404, L60}

\bibitem[\protect\citeauthoryear{{Salpeter}}{{Salpeter}}{1955}]{Salpeter55}
{Salpeter} E.~E.,  1955, \mn@doi [\apj] {10.1086/145971}, \href
  {http://adsabs.harvard.edu/abs/1955ApJ...121..161S} {121, 161}

\bibitem[\protect\citeauthoryear{{Sanders} \& {McGaugh}}{{Sanders} \&
  {McGaugh}}{2002}]{Sanders&McGaugh02}
{Sanders} R.~H.,  {McGaugh} S.~S.,  2002, \mn@doi [\araa]
  {10.1146/annurev.astro.40.060401.093923}, \href
  {http://adsabs.harvard.edu/abs/2002ARA%26A..40..263S} {40, 263}

\bibitem[\protect\citeauthoryear{{Sault}, {Teuben}  \& {Wright}}{{Sault}
  et~al.}{1995}]{Sault+95}
{Sault} R.~J.,  {Teuben} P.~J.,   {Wright} M.~C.~H.,  1995, in {Shaw} R.~A.,
  {Payne} H.~E.,   {Hayes} J.~J.~E.,  eds,  Astronomical Society of the Pacific
  Conference Series Vol. 77, Astronomical Data Analysis Software and Systems
  IV. p.~433 (\mn@eprint {} {astro-ph/0612759})

\bibitem[\protect\citeauthoryear{{Sault}, {Staveley-Smith}  \& {Brouw}}{{Sault}
  et~al.}{1996}]{Sault+96}
{Sault} R.~J.,  {Staveley-Smith} L.,   {Brouw} W.~N.,  1996, \aaps, \href
  {http://adsabs.harvard.edu/abs/1996A%26AS..120..375S} {120, 375}

\bibitem[\protect\citeauthoryear{{Skibba} et~al.,}{{Skibba}
  et~al.}{2012}]{Skibba+12}
{Skibba} R.~A.,  et~al., 2012, \mn@doi [\apj] {10.1088/0004-637X/761/1/42},
  \href {http://adsabs.harvard.edu/abs/2012ApJ...761...42S} {761, 42}

\bibitem[\protect\citeauthoryear{{Stanimirovi{\'c}}, {Staveley-Smith},
  {Dickey}, {Sault}  \& {Snowden}}{{Stanimirovi{\'c}}
  et~al.}{1999}]{Stanimirovic+99}
{Stanimirovi{\'c}} S.,  {Staveley-Smith} L.,  {Dickey} J.~M.,  {Sault} R.~J.,
  {Snowden} S.~L.,  1999, \mn@doi [\mnras] {10.1046/j.1365-8711.1999.02013.x},
  \href {http://adsabs.harvard.edu/abs/1999MNRAS.302..417S} {302, 417}

\bibitem[\protect\citeauthoryear{{Stanimirovi{\'c}}, {Staveley-Smith}  \&
  {Jones}}{{Stanimirovi{\'c}} et~al.}{2004}]{Stanimirovic+04}
{Stanimirovi{\'c}} S.,  {Staveley-Smith} L.,   {Jones} P.~A.,  2004, \mn@doi
  [\apj] {10.1086/381869}, \href
  {http://adsabs.harvard.edu/abs/2004ApJ...604..176S} {604, 176}

\bibitem[\protect\citeauthoryear{{Staveley-Smith}, {Sault}, {Hatzidimitriou},
  {Kesteven}  \& {McConnell}}{{Staveley-Smith}
  et~al.}{1997}]{Staveley-Smith+97}
{Staveley-Smith} L.,  {Sault} R.~J.,  {Hatzidimitriou} D.,  {Kesteven} M.~J.,
  {McConnell} D.,  1997, \mn@doi [\mnras] {10.1093/mnras/289.2.225}, \href
  {http://adsabs.harvard.edu/abs/1997MNRAS.289..225S} {289, 225}

\bibitem[\protect\citeauthoryear{{Steer}, {Dewdney}  \& {Ito}}{{Steer}
  et~al.}{1984}]{Steer+84}
{Steer} D.~G.,  {Dewdney} P.~E.,   {Ito} M.~R.,  1984, \aap, \href
  {http://adsabs.harvard.edu/abs/1984A%26A...137..159S} {137, 159}

\bibitem[\protect\citeauthoryear{{Subramanian} \& {Subramaniam}}{{Subramanian}
  \& {Subramaniam}}{2012}]{Subramanian&Subramanian12}
{Subramanian} S.,  {Subramaniam} A.,  2012, \mn@doi [\apj]
  {10.1088/0004-637X/744/2/128}, \href
  {http://adsabs.harvard.edu/abs/2012ApJ...744..128S} {744, 128}

\bibitem[\protect\citeauthoryear{{Subramanian} \& {Subramaniam}}{{Subramanian}
  \& {Subramaniam}}{2015}]{Subramanian&Subramanian15}
{Subramanian} S.,  {Subramaniam} A.,  2015, \mn@doi [\aap]
  {10.1051/0004-6361/201424248}, \href
  {http://adsabs.harvard.edu/abs/2015A%26A...573A.135S} {573, A135}

\bibitem[\protect\citeauthoryear{{Swaters}}{{Swaters}}{1999}]{Swaters99}
{Swaters} R.~A.,  1999, PhD thesis, , Rijksuniversiteit Groningen, (1999)

\bibitem[\protect\citeauthoryear{{Swaters}, {Sanders}  \& {McGaugh}}{{Swaters}
  et~al.}{2010}]{Swaters+10}
{Swaters} R.~A.,  {Sanders} R.~H.,   {McGaugh} S.~S.,  2010, \mn@doi [\apj]
  {10.1088/0004-637X/718/1/380}, \href
  {http://adsabs.harvard.edu/abs/2010ApJ...718..380S} {718, 380}

\bibitem[\protect\citeauthoryear{{Vazdekis}, {S{\'a}nchez-Bl{\'a}zquez},
  {Falc{\'o}n-Barroso}, {Cenarro}, {Beasley}, {Cardiel}, {Gorgas}  \&
  {Peletier}}{{Vazdekis} et~al.}{2010}]{Vazdekis+10}
{Vazdekis} A.,  {S{\'a}nchez-Bl{\'a}zquez} P.,  {Falc{\'o}n-Barroso} J.,
  {Cenarro} A.~J.,  {Beasley} M.~A.,  {Cardiel} N.,  {Gorgas} J.,   {Peletier}
  R.~F.,  2010, \mn@doi [\mnras] {10.1111/j.1365-2966.2010.16407.x}, \href
  {http://adsabs.harvard.edu/abs/2010MNRAS.404.1639V} {404, 1639}

\bibitem[\protect\citeauthoryear{{Vieira} et~al.,}{{Vieira}
  et~al.}{2010}]{Vieira+09}
{Vieira} K.,  et~al., 2010, \mn@doi [\aj] {10.1088/0004-6256/140/6/1934}, \href
  {http://adsabs.harvard.edu/abs/2010AJ....140.1934V} {140, 1934}

\bibitem[\protect\citeauthoryear{{Werner} et~al.,}{{Werner}
  et~al.}{2004}]{Werner+04}
{Werner} M.~W.,  et~al., 2004, \mn@doi [\apjs] {10.1086/422992}, \href
  {http://adsabs.harvard.edu/abs/2004ApJS..154....1W} {154, 1}

\bibitem[\protect\citeauthoryear{{Zibetti}, {Charlot}  \& {Rix}}{{Zibetti}
  et~al.}{2009}]{Zibetti+09}
{Zibetti} S.,  {Charlot} S.,   {Rix} H.-W.,  2009, \mn@doi [\mnras]
  {10.1111/j.1365-2966.2009.15528.x}, \href
  {http://adsabs.harvard.edu/abs/2009MNRAS.400.1181Z} {400, 1181}

\bibitem[\protect\citeauthoryear{{de Blok} \& {McGaugh}}{{de Blok} \&
  {McGaugh}}{1998}]{deBlok&McGaugh98}
{de Blok} W.~J.~G.,  {McGaugh} S.~S.,  1998, \mn@doi [\apj] {10.1086/306390},
  \href {http://adsabs.harvard.edu/abs/1998ApJ...508..132D} {508, 132}

\bibitem[\protect\citeauthoryear{{de Blok}, {Walter}, {Brinks}, {Trachternach},
  {Oh}  \& {Kennicutt}}{{de Blok} et~al.}{2008}]{deBlok+08}
{de Blok} W.~J.~G.,  {Walter} F.,  {Brinks} E.,  {Trachternach} C.,  {Oh}
  S.-H.,   {Kennicutt} Jr. R.~C.,  2008, \mn@doi [\aj]
  {10.1088/0004-6256/136/6/2648}, \href
  {http://adsabs.harvard.edu/abs/2008AJ....136.2648D} {136, 2648}

\bibitem[\protect\citeauthoryear{{de Vaucouleurs}}{{de
  Vaucouleurs}}{1955}]{deVaucouleurs55}
{de Vaucouleurs} G.,  1955, \mn@doi [\aj] {10.1086/107218}, \href
  {http://adsabs.harvard.edu/abs/1955AJ.....60..219D} {60, 219}

\bibitem[\protect\citeauthoryear{{van Albada}, {Bahcall}, {Begeman}  \&
  {Sancisi}}{{van Albada} et~al.}{1985}]{vanAlbada+85}
{van Albada} T.~S.,  {Bahcall} J.~N.,  {Begeman} K.,   {Sancisi} R.,  1985,
  \mn@doi [\apj] {10.1086/163375}, \href
  {http://adsabs.harvard.edu/abs/1985ApJ...295..305V} {295, 305}

\bibitem[\protect\citeauthoryear{{van der Marel} \& {Cioni}}{{van der Marel} \&
  {Cioni}}{2001}]{vdM&Cioni01}
{van der Marel} R.~P.,  {Cioni} M.-R.~L.,  2001, \mn@doi [\aj]
  {10.1086/323099}, \href {http://adsabs.harvard.edu/abs/2001AJ....122.1807V}
  {122, 1807}

\bibitem[\protect\citeauthoryear{{van der Marel} \& {Sahlmann}}{{van der Marel}
  \& {Sahlmann}}{2016}]{vanderMarel+16}
{van der Marel} R.~P.,  {Sahlmann} J.,  2016, \mn@doi [\apjl]
  {10.3847/2041-8205/832/2/L23}, \href
  {http://adsabs.harvard.edu/abs/2016ApJ...832L..23V} {832, L23}

\bibitem[\protect\citeauthoryear{{van der Marel}, {Alves}, {Hardy}  \&
  {Suntzeff}}{{van der Marel} et~al.}{2002}]{vdM+02}
{van der Marel} R.~P.,  {Alves} D.~R.,  {Hardy} E.,   {Suntzeff} N.~B.,  2002,
  \mn@doi [\aj] {10.1086/343775}, \href
  {http://adsabs.harvard.edu/abs/2002AJ....124.2639V} {124, 2639}

\makeatother
\end{thebibliography}

\label{lastpage}
\end{document}